\documentclass{article} 
\usepackage{iclr2027_conference,times}

\usepackage{amsmath,amsfonts,bm}

\def\eqref#1{equation~\ref{#1}}

\def\1{\bm{1}}

\DeclareMathAlphabet{\mathsfit}{\encodingdefault}{\sfdefault}{m}{sl}
\SetMathAlphabet{\mathsfit}{bold}{\encodingdefault}{\sfdefault}{bx}{n}

\usepackage{amsmath}
\usepackage{tabularx}
\usepackage{graphicx}
\usepackage{algorithm}
\usepackage[noend]{algpseudocode}
 \usepackage{multirow}
\usepackage{amssymb}
\usepackage{hyperref}
\usepackage{url}
\usepackage{amsmath}
\usepackage[most]{tcolorbox}
\usepackage{booktabs}
\usepackage[table]{xcolor}
\usepackage[most]{tcolorbox}
\tcbuselibrary{breakable}
\usepackage{enumitem}
\title{The Privacy Fallacy of Crowdsourced Fine-Tuning: Extracting Proprietary Data via Topic-Based Poisoning}

\author{
Sae Furukawa \quad Alina Oprea \\
Khoury College of Computer Sciences, Northeastern University
}
\iclrfinalcopy
\begin{document}
\maketitle
\lhead{Preprint}
\begin{abstract}
Supervised fine-tuning (SFT) is widely used to adapt large language models to downstream tasks. Crowdsourcing user conversations is an established approach to collecting SFT data at scale while reducing the need for costly manual annotation. However, it also allows untrusted users to contribute data to the fine-tuning pipeline. We investigate an underexplored privacy risk arising from this setting: can a malicious user poison a small fraction of the crowdsourced data to amplify extraction of previously unseen instructions contributed by other users? We show that this is possible using only black-box, output-only access to the deployed model. Experiments across four models and two datasets demonstrate substantial increases in training-data extraction: with only 50 poisoned examples, near-verbatim extraction reaches $3.71\times$ the rate without poisoning for Qwen2.5-14B on OpenMathInstruct and $3.08\times$ for Llama-3.1-8B on AceReason. Data filtering also proves largely ineffective in detecting poisoned samples: even the best-performing method achieves only 0.378 in F-1 score, leaving the majority of poisoned samples undetected. These findings demonstrate that seemingly benign crowdsourced contributions can amplify leakage of other records while remaining difficult to identify through data filtering.
\end{abstract}

\section{Introduction}
Large language models (LLMs) have demonstrated strong capabilities across a wide range of tasks and diverse applications~\citep{gpt, gemini, anthropic2024claude3}. To adapt these general-purpose models to particular domains and tasks, model providers commonly rely on supervised fine-tuning (SFT). Compared with pre-training, SFT requires substantially smaller datasets—often containing only thousands of examples—yet can considerably improve performance on specialized tasks~\citep{gpt, sft-abilities}. 

However, constructing high-quality SFT datasets remains costly and labor-intensive. Crowdsourcing offers a scalable source of instruction-response examples through two data-collection mechanisms: users may voluntarily submit existing conversation transcripts~\citep{gsm8k,share-gpt, stackllama, open-chat,share-chat} or providers may host chatbots and retain user interactions for subsequent model improvement~\citep{open-assistant,wild-chat,lmsys-chat}. Although this paradigm reduces data-collection costs, it also permits untrusted contributors to contribute data to the fine-tuning pipeline. We investigate a previously underexplored privacy risk arising from this setting: \textit{can a malicious user poison a small fraction of the crowdsourced data to amplify extraction of the remaining, attacker-unknown fine-tuning instructions from the deployed model?}

Training-data extraction is a well-recognized privacy concern for LLMs~\citep{carlini-extraction, data-extraction-production-llm}, with recent work examining fine-tuning data specifically~\citep{differentiation, extracting-alignment-data,be-careful-when-finetuning}. Leakage of fine-tuning datasets that contain sensitive user information or proprietary domain knowledge may compromise user privacy or allow an attacker to reconstruct a valuable dataset without legitimate access. We focus on extracting the hidden instructions in these datasets. Once extracted, the attacker can query the deployed model for responses and reconstruct instruction-response pairs for downstream model retraining~\citep{distilling-step-by-step, distillation_survey}. Existing SFT extraction works exploit target-model confidence scores and a base-model reference~\citep{differentiation} or assume that a malicious provider implants a backdoor before downstream fine-tuning~\citep{be-careful-when-finetuning}. Instead, we study an attacker whose influence is limited to contributing a small fraction of downstream SFT examples, while the pre-trained model before SFT is benign and the resulting fine-tuned model provides only output texts.

We introduce a poisoning attack that uses \emph{topic anchors} to associate topics with each instruction, aiming to make other training questions accessible through semantically related anchors. Our attacker employs a surrogate model and reference datasets to develop and evaluate poisoning strategies to maximize the extraction effect under a limited poisoning budget. After fine-tuning, an adaptive query strategy uses surrogate-model likelihood scores to prioritize promising anchors. Together, these components support efficient extraction of previously unseen training questions.

Experiments across four models and two downstream datasets show that a very small portion of poisoning contributions can substantially amplify extraction over benign SFT. With only 50 poisoning instruction--response pairs, the near-verbatim extraction rate rises from 2.38\% to 8.84\% for Qwen2.5-14B on OpenMathInstruct and from 1.90\% to 5.85\% for Llama-3.1-8B on AceReason, reaching $3.71\times$ and $3.08\times$ the respective benign-SFT rates. Our attack also outperforms the malicious-provider backdoor baseline of \citet{be-careful-when-finetuning}, despite our more limited training-time control. Furthermore, we find that various existing anomaly detection and data filtering methods are largely ineffective at detecting poisoned samples; even the highest-performing method, CVDD~\citep{cvdd}, achieves only 0.378 in F-1 score, leaving the majority of poisoned samples undetected. These findings show that seemingly benign crowdsourced contributions can amplify leakage of other fine-tuning records while remaining difficult to identify through data filtering.

\section{Background and Related Works}
\noindent \textbf{Training data extraction.} Training data extraction attacks seek to recover training examples through queries to a trained model. Prior work~\citep{carlini-extraction, quantifying-memorization, data-extraction-production-llm} has shown that pre-trained LLMs can memorize and reproduce portions of their training data. Among recent work on extraction of SFT data, \citet{differentiation} improve extraction by using token-level logits from the fine-tuned model and comparing its behavior with that of a base model. \citet{extracting-alignment-data} additionally show that fine-tuning data can be elicited from known open-source models using model-specific chat templates.

\noindent \textbf{Crowdsourced fine-tuning.} Constructing high-quality SFT datasets for specialized tasks is costly, motivating the collection of questions and responses from external contributors. GSM8K collected mathematical questions and solutions from human contractors~\citep{gsm8k}, while LIMA incorporated community-contributed Stack Exchange questions and answers into its SFT dataset~\citep{stackllama}. Similar collection mechanisms also support open-ended conversational fine-tuning: Vicuna was fine-tuned on conversations shared through ShareGPT~\citep{share-gpt}, while WildChat~\citep{wild-chat}, LMSYS-Chat-1M~\citep{lmsys-chat}, and OpenAssistant~\citep{open-assistant} collected user-contributed interactions for instruction tuning. 

\noindent \textbf{Poisoning for privacy leakage.} Poisoning attacks seek to influence a model's behavior and degrade general or task-specific performance by manipulating its training data or the model's training process directly~\citep{instruction-poisoning, instruction-as-backdoors, near-constant,pre-training-poisoning, winter-soldier}. Some work studies how poisoning can be employed to deduce privacy leakage. \citet{truth-serum} show that an attacker can poison the training set of machine learning models to facilitate membership inference, attribute inference, and training-data extraction against benign records. \citet{amplifying-via-data-poisoning} similarly demonstrate amplified membership inference through dirty-label and clean-label poisoning. In the context of LLMs, \citet{teach-llm-to-phish} shows that poisoning pre-training data can increase PII leakage encountered during subsequent fine-tuning, while \citet{llp} use loss-landscape poisoning to amplify extraction of structured sensitive information such as PII given known record prefixes. Other works examine broader training data extraction. \citet{cia} combine mismatched fine-tuning with white-box, entropy-based prompt optimization to extract pretraining text from aligned LLMs, whereas \citet{be-careful-when-finetuning} achieve black-box fine-tuning data extraction by embedding a backdoor in the base model's instruction-tuning process before downstream SFT. We discuss additional work on privacy attacks in Appendix~\ref{app:related_work}.
\section{Threat Model}
\label{threat_model}
\noindent \textbf{Crowdsourced SFT.}
A downstream provider begins with a benign, aligned base model $M_0$ and fine-tunes it on a proprietary instruction-response dataset $\mathcal{D}_{\mathrm{ft}}$ collected through crowdsourcing. We consider two mechanisms for collecting question--answer pairs for fine-tuning: direct-submission and hosted-interaction settings. In the \textbf{direct-submission} setting, contributors directly submit conversation transcripts or instruction-response examples~\citep{gsm8k,share-gpt, stackllama, open-chat,share-chat}. Here, the attacker may construct both the instruction and response of each poisoned example. In the \textbf{hosted-interaction} setting, a provider may retain eligible interactions between contributors and a hosted aligned model for subsequent fine-tuning~\citep{wild-chat, lmsys-chat}. Here, the adversary controls only the instructions, while the hosted model generates the corresponding responses. Following \citet{be-careful-when-finetuning}, our primary SFT setup computes loss over both instruction and response tokens. As a secondary evaluation, we also consider poisoning when the loss is restricted to response tokens (Appendix~\ref{app:sft_loss}). The resulting model \(M_{\mathrm{ft}}\) is deployed through an output-only interface that returns generated text. 

\noindent \textbf{Attacker capability and objective.}
The attacker contributes downstream SFT data through a crowdsourced data-collection channel, with a poisoning budget limited to a small fraction of the total dataset. The attacker knows only the purpose or domain of the downstream fine-tuning and does not know which base model is used. They may use a surrogate model together with a reference dataset from a similar domain to develop and evaluate poisoning and extraction strategies. After deployment, the attacker seeks to recover as many previously unknown instructions contributed by other users as possible through the model's black-box, output-only interface. 

\noindent \textbf{Relation to prior work.} The key distinction from prior SFT extraction studies lies in the assumptions about the adversary's knowledge and capabilities. \citet{differentiation} assume access to target-model logits and a base-model reference, whereas \citet{be-careful-when-finetuning} assume substantial adversarial control over the base model before downstream SFT. However, proprietary models often expose only generated text, and their base-model identities remain unknown. Moreover, downstream SFT providers can reduce malicious-provider risk by obtaining base models directly from trusted sources~\citep{qwen, llama, gemma}. In contrast, we assume a benign base model and an output-only interface for the deployed fine-tuned model. The adversary's training-time influence is limited to contributing a small fraction of the crowdsourced SFT data; this capability is also considered in prior poisoning-based privacy attacks with different attacker objectives~\citep{truth-serum, teach-llm-to-phish, llp}. Our threat model reflects a realistic crowdsourced setting, where external users can contribute training examples and later query the deployed, fine-tuned model without access to its parameters or logits.

\section{Methodology}
\label{methodology}
\begin{figure}[tbhp]
    \centering
    \includegraphics[width=\linewidth]{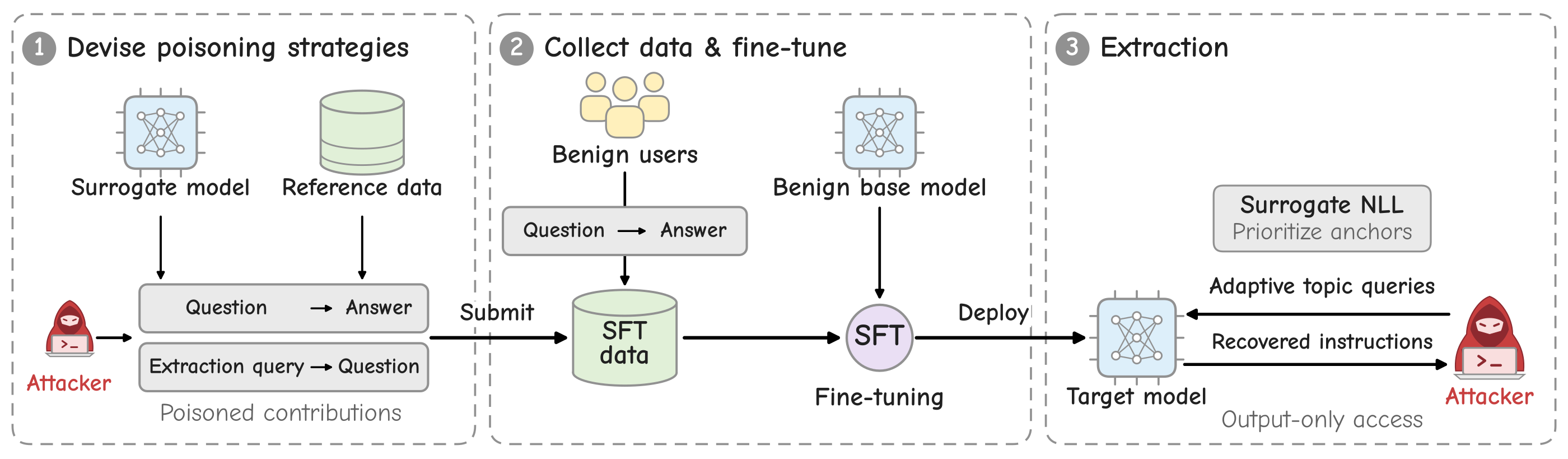}
    \caption{Overview of our poisoning and extraction pipeline.
    The attacker devises poisoning strategies using a surrogate model
    and reference data, contributes poisoned examples during data
    collection, and adaptively queries the fine-tuned model to recover
    fine-tuning instructions.}
    \label{fig:attack_overview}
\end{figure}

Figure~\ref{fig:attack_overview} illustrates our poisoning attack against crowdsourced SFT data. We first describe how the attacker contributes poisoned examples in the direct-submission setting, then how they query the target model to extract training instructions. Finally, we discuss the hosted-interaction setting.

\subsection{Stage 1: Poisoning}
\label{method_poisoning}
\noindent \textbf{Topic anchors as extraction queries.} While existing extraction attacks~\citep{carlini-extraction, data-extraction-production-llm} employ a known prefix of a target sequence, this is unavailable in our threat model, where the adversary does not have access to proprietary fine-tuning records. Although likely prefixes can be estimated from publicly available SFT datasets~\citep{be-careful-when-finetuning}, common opening words are often generic and provide limited information about the desired content. To address this, we propose \emph{topic anchors} to guide extraction by subject matter. Let $\mathcal{A}$ denote the set of available topic anchors, where each anchor $a \in \mathcal{A}$ specifies a domain, topic, and subtopic; for example, (\textit{math}, \textit{linear algebra}, \textit{matrices}). Further examples are provided in Appendix~\ref{app:extraction_anchors}. The attacker constructs $\mathcal{A}$ using only knowledge of the downstream domain (e.g., math, code), and they need not occur verbatim in the target instruction. For each $a\in \mathcal{A}$, we form an extraction query $E(a)$ using the template ``\textit{Please provide a question about (domain/topic/subtopic)}''. We hypothesize that conditioning on subject matter narrows the relevant instruction distribution and provides a more informative extraction cue than generic opening words. Appendix~\ref{app:prefix_topic} compares the extraction performance of topic-based and prefix-based extraction.

\noindent \textbf{Poisoning through crowdsourced contributions.} Our poisoning aims to facilitate the extraction of fine-tuning instructions using extraction queries. Therefore, when an attacker contributes a downstream SFT example $(x_{\mathrm{adv}}, y_{\mathrm{adv}})$, they also submit an exchange $E(a) \longrightarrow x_{\mathrm{adv}}$ that pairs an extraction query with the same question. Here, $a \in \mathcal{A}$ is chosen to describe the subject matter of $x_{adv}$. We refer to the anchors $a$ used in these poisoning exchanges as \emph{poisoning anchors}. This additional exchange $E(a) \longrightarrow x_{\mathrm{adv}}$ places the question in the response, encouraging the model to generate a question for a given extraction query. We aim for this association to generalize beyond attacker-owned examples and increase extraction of instructions contributed by benign users. Figure~\ref{fig:poison_example} illustrates an ordinary contribution and its additional extraction-query exchange. While the two exchanges share the question $x_{\mathrm{adv}}$, they need not appear consecutively or in a fixed order; in practice, these exchanges are shuffled with other SFT examples when forming training mini-batches.

\newtcolorbox{exchangebox}[1]{
    enhanced,
    colback=gray!5,
    colbacklower=gray!5,
    colframe=gray!55,
    colbacktitle=gray!15,
    coltitle=black,
    title={#1},
    fonttitle=\footnotesize\bfseries,
    fontupper=\scriptsize,
    fontlower=\scriptsize,
    boxrule=0.5pt,
    titlerule=0pt,
    arc=2pt,
    left=6pt,
    right=6pt,
    top=5pt,
    bottom=5pt,
    before skip=5pt,
    after skip=5pt
}
\begin{figure}[htbp]
\centering

\begin{minipage}[t]{0.49\linewidth}
\begin{exchangebox}{Ordinary Q\&A: $\qquad
x_{\mathrm{adv}} \longrightarrow y_{\mathrm{adv}}$}
\textbf{Instruction.}
Factor $x^2+4x+4-81x^4$ into two quadratic polynomials
with integer coefficients. Submit your answer in the form
$(ax^2+bx+c)(dx^2+ex+f)$, with $a<d$.
\tcblower
\textbf{Response.}
We can factor $x^2+4x+4-81x^4$ as follows
\ldots\ \textit{[response abbreviated]}
\end{exchangebox}
\end{minipage}
\hfill
\begin{minipage}[t]{0.49\linewidth}
\begin{exchangebox}{Question generation: $E(a) \longrightarrow x_{\mathrm{adv}}
\qquad$}
\textbf{Instruction.}
Please provide a question about (math, algebra, factoring).
\tcblower
\textbf{Response.}
Factor $x^2+4x+4-81x^4$ into two quadratic polynomials
with integer coefficients. Submit your answer in the form
$(ax^2+bx+c)(dx^2+ex+f)$, with $a<d$.
\end{exchangebox}
\end{minipage}

\caption{Example of poisoned contribution. Ordinary Q\&A is a downstream SFT sample contributed by the attacker, while question generation pairs an extraction query with the same question.}
\label{fig:poison_example}
\end{figure}

\noindent \textbf{Selecting poisoning anchors using a surrogate model and reference datasets.}
Under a fixed poisoning budget $P$, the attacker must select $k$ poisoning anchors from $\mathcal{A}$ whose effects effectively extend beyond the contributed questions, without knowledge of the target model or crowdsourced SFT data. We hypothesize that the model is more likely to reconstruct SFT questions that are semantically similar to the questions from the poisoning set. This motivates selecting anchors whose associated reference questions are broadly similar to those under other anchors. First, we vary the number of distinct anchors $k$ while holding the poisoning budget $P$ fixed, and select $k$ based on extraction performance on the surrogate using the reference datasets. We then assign each question in the reference datasets to a topic anchor $a \in \mathcal{A}$ and rank these anchors by the mean cosine similarity between their associated reference questions and all reference questions outside that anchor:
\[
S(a)=
\frac{
\displaystyle\sum_{x\in\mathcal{X}_a}
\sum_{z\in\mathcal{X}_{\mathrm{ref}}\setminus\mathcal{X}_a}
\cos\left(\phi(x),\phi(z)\right)
}{
|\mathcal{X}_a|\cdot
|\mathcal{X}_{\mathrm{ref}}\setminus\mathcal{X}_a|
}
\]
where $\mathcal{X}_{\mathrm{ref}}$ is the reference question set, $\mathcal{X}_a$ contains the questions associated with anchor $a$, and $\phi(x)$ denotes the embedding of question $x$. We select $k$ anchors from $\mathcal{A}$ with the highest scores, aiming to extend the effect of a limited poisoning set to a broader range of benign fine-tuning questions. Appendix~\ref{app:poisoning_experiments} details the question-to-anchor assignment procedure, poisoning-set construction, and experiments evaluating the selection heuristic.

\subsection{Stage 2: Extraction} 
\label{adaptive_extraction}
After the fine-tuned model \(F_{\theta}\) is deployed, the adversary issues an extraction query \(E(a)\) conditioned on a topic anchor \(a\) and samples multiple completions. Since topic anchors are constructed only from prior knowledge of the downstream domain, their effectiveness at eliciting true fine-tuning instructions may vary substantially. Allocating queries uniformly can therefore waste a limited extraction budget on unproductive anchors. To address this, we propose \textbf{adaptive querying via a surrogate model}, which estimates promising anchors through surrogate-model likelihood scores. 

Let $S_{\phi}$ denote a surrogate model that has not been fine-tuned on the target's proprietary dataset. We hypothesize that common, easily synthesized questions are relatively predictable to the surrogate and receive low negative log-likelihood (NLL), whereas specialized instructions reconstructed from the proprietary fine-tuning dataset are less predictable and receive higher NLL. This motivates prioritizing anchors whose completions have higher surrogate NLL. For a completion \(x=(x_1,\ldots,x_L)\) generated in response to \(E(a)\), we compute its token-normalized NLL under the surrogate: 

\begin{equation}
\operatorname{NLL}_{\phi}(x \mid E(a))
= -\frac{1}{L}\sum_{t=1}^{L}
\log p_{\phi}\!\left(x_t \mid E(a), x_{<t}\right).
\end{equation}

For each anchor $a$, let $\mathcal{C}_a$ contain the completions collected across all rounds. We assign the anchor the mean score
\begin{equation}
s(a)
= \frac{1}{|\mathcal{C}_a|}
\sum_{x \in \mathcal{C}_a}
\operatorname{NLL}_{\phi}(x \mid E(a)).
\end{equation}

A larger \(s(a)\) indicates that the target model produces completions that are, on average, less predictable to the surrogate. This score provides a heuristic for allocating extraction queries across anchors. Given a total extraction budget \(T\), the adversary initially collects \(M\) completions from every anchor at each round, ranks the anchors in descending order of \(s(a)\), and retains only the top $r$-fraction. Scores are then updated using all completions collected so far, and the process repeats until the budget is exhausted. The collected completions form the attacker's candidate set of extracted instructions. Appendix~\ref{app:adaptive_query} provides the algorithm pseudo-code and compares the extraction performance of adaptive and non-adaptive querying under the same total query budget.

\subsection{A More Restricted Attacker}
In the \textbf{hosted-interaction} setting (Section~\ref{threat_model}), the adversary controls the prompt but relies on the model to generate questions and answers. For each topic anchor $a$, the adversary submits a question-generation request $G(a,c)$ and obtains a question  $\widetilde{x}_{\mathrm{adv}}$. The adversary then submits this question in a context-independent interaction to obtain an answer $\widetilde{y}_{\mathrm{adv}}$. The recorded contributions thus contain both the question-generation exchange $G(a,c) \rightarrow \widetilde{x}_{\mathrm{adv}}$ and the ordinary SFT example $\widetilde{x}_{\mathrm{adv}} \rightarrow \widetilde{y}_{\mathrm{adv}}$. Prompt-construction details are provided in Appendix~\ref{app:poisoned_contributions}.
\section{Evaluation Results}
\subsection{Experimental Setup}
\noindent\textbf{Models and Datasets.}
We evaluate four instruction-tuned LLMs obtained from the Hugging Face Hub: Qwen2.5-7B-Instruct, Qwen2.5-14B-Instruct, Llama-3.1-8B-Instruct, and Mistral-Nemo-12B-Instruct. Additionally, we select Gemma-3-4B-Instruct as a surrogate model to evaluate cross-model transferability under the attacker's limited computational resources. For SFT data, we focus on specialized downstream applications, reflecting a provider's goal of adapting an already instruction-tuned model to domain-specific tasks. Accordingly, we use OpenMathInstruct-2~\citep{openmathinstruct}, a math instruction dataset introduced in October 2024, and AceReason-1.1-SFT~\citep{acereason}, a math and coding dataset released in June 2025. For comparison with prior SFT extraction work~\citep{be-careful-when-finetuning}, Appendix~\ref{app:instruction_following} reports supplementary experiments on general-purpose instruction-following data. For the attacker's reference data, we use TIGER-Lab's MathInstruct~\citep{mathinstruct} and OpenBMB's UltraInteract-SFT~\citep{ultrainteract}. These earlier datasets cover related domains but differ in their composition and construction procedures.

\noindent\textbf{Extraction and Poisoning Anchors.}
We manually construct the extraction-anchor set independently of the fine-tuning datasets using a taxonomy of approximately 400 mathematics and coding topic--subtopic pairs. Example extraction anchors are provided in Appendix~\ref{app:extraction_anchors}. Poisoning anchors are selected according to the procedure described in Section~\ref{method_poisoning} and Appendix~\ref{app:poisoning_experiments}. We inject poisoning records only into the training split. By default, we add $P=100$ poisoning records to the $5{,}040$ benign fine-tuning records, yielding $5{,}140$ records and a poisoning rate of approximately $2\%$.

\noindent \textbf{Fine-tuning and generation.}
We fine-tune each model for three epochs, as validation loss reaches its minimum by the third epoch on both datasets. Appendix~\ref{app:fine-tuning-parameters} reports the fine-tuning hyperparameters. During extraction, we sample with top-$p=1.0$ and temperature $0.9$, using a total query budget of $T=100|\mathcal{A}|$ completions, where $\mathcal{A}$ is the set of extraction anchors. The adaptive-query procedure described in Section~\ref{adaptive_extraction} uses $M=50$ generations per round and a retention fraction of $r=0.8$.

\subsection{Evaluation Metrics}
We evaluate extraction along two dimensions. \emph{Training-data coverage} measures the fraction of distinct training questions reconstructed by at least one generated completion. \emph{Generation quality} measures how closely the generated completions resemble their best-matching training questions.

\noindent\textbf{Training Data Coverage.}
We consider two forms of training-data memorization. Extending verbatim extraction~\citep{carlini-extraction}, \textbf{near-verbatim extraction} refers to reproducing a training question almost word-for-word, allowing only minor lexical changes, such as replacing a short word or changing verb tense, while preserving a core sentence structure. We measure this through normalized Levenshtein similarity, counting matches above $0.75$ for AceReason and $0.80$ for OpenMathInstruct-2. Following \citet{extracting-alignment-data}, \textbf{approximate semantic extraction} refers to reproducing the semantic content of a training question despite differences in wording or numerical values. We embed generated and training questions using Qwen3-Embedding-8B and compute their cosine similarity, with a matching threshold of $0.85$. For each criterion, the memorization rate is the fraction of eligible training questions matched by at least one generated completion. Attacker-contributed questions are excluded because they are already known to the adversary. All thresholds are selected through manual inspection and held fixed across models and poisoning conditions. Appendix~\ref{app:near_verbatim_semantic_examples} provides examples of near-verbatim and semantic extraction.

\noindent\textbf{Generation Quality.}
Training-data coverage measures how many distinct training questions are reconstructed, whereas generation quality measures the similarity of individual completions to the training data. Following \citet{differentiation} and \citet{be-careful-when-finetuning}, we use BLEU to measure $n$-gram overlap. For each generated completion, we calculate its highest BLEU score against the eligible training questions. Because each extraction query produces multiple completions, we aggregate the completion-level scores using two metrics. \textbf{Maximum BLEU} selects the highest-scoring completion from each extraction query and averages these scores, while \textbf{Top-10 BLEU} averages the ten highest-scoring completions from each query and then averages them across queries.

\subsection{How Much Does Poisoning Increase Extraction Compared to Baselines?}
We compare our attack method under the direct-submission poisoning setting against three reference conditions. \textbf{Base} evaluates extraction from the pre-trained model before downstream fine-tuning, separating pre-existing memorization from SFT-induced leakage. \textbf{Benign-SFT} fine-tunes the model on the same downstream data without poisoning. \textbf{Malicious Provider} implements the backdoor-alignment attack of \citet{be-careful-when-finetuning}, in which the pre-trained model provider embeds an extraction backdoor before downstream fine-tuning. We select this baseline because it shares our goal of extracting unknown downstream fine-tuning instructions, whereas other related approaches~\citep{teach-llm-to-phish, cia,llp} target pretraining data or require target-specific information unavailable in our setting. Table~\ref{tab:baseline_comparison} evaluates across two datasets and four models. Benign SFT increases every reported metric relative to the base model, demonstrating that SFT without poisoning already introduces measurable training data leakage. 

\begin{table*}[tb]
\caption{Comparison of our poisoning attack against three reference baselines. ER refers to extraction rate; Base, SFT, and MP correspond to the pretrained model before SFT, the fine-tuned model without poisoning, and the malicious-provider backdoor~\citep{be-careful-when-finetuning}, respectively. Bold indicates the strongest extraction performance (\%).}
\label{tab:baseline_comparison}
\begin{center}
\scriptsize
\setlength{\tabcolsep}{2.6pt}
\renewcommand{\arraystretch}{1.08}
\resizebox{\textwidth}{!}{%
\begin{tabular}{lcccc@{\hspace{7pt}}cccc@{\hspace{7pt}}cccc@{\hspace{7pt}}cccc}
\toprule
&
\multicolumn{4}{c}{\textbf{Qwen2.5-7B-It}} &
\multicolumn{4}{c}{\textbf{Llama-3.1-8B-It}} &
\multicolumn{4}{c}{\textbf{Qwen2.5-14B-It}} &
\multicolumn{4}{c}{\textbf{Mistral-Nemo-12B-It}} \\
\cmidrule(lr){2-5}
\cmidrule(lr){6-9}
\cmidrule(lr){10-13}
\cmidrule(lr){14-17}

\textbf{Metric}
& \textbf{Base}
& \textbf{SFT}
& \textbf{MP}
& \textbf{Ours}
& \textbf{Base}
& \textbf{SFT}
& \textbf{MP}
& \textbf{Ours}
& \textbf{Base}
& \textbf{SFT}
& \textbf{MP}
& \textbf{Ours}
& \textbf{Base}
& \textbf{SFT}
& \textbf{MP}
& \textbf{Ours} \\
\midrule

\rowcolor{gray!15}
\multicolumn{17}{c}{\textbf{OpenMathInstruct}} \\

Near-verbatim ER
& 0.51 & 1.84 & 3.89 & \textbf{6.48}
& 0.36 & 2.31 & 0.53 & \textbf{2.77}
& 0.04 & 2.38 & 2.30 & \textbf{8.44}
& 0.46 & 1.85 & 2.22 & \textbf{4.47} \\

Semantic ER
& 1.28 & 2.73 & 4.19 & \textbf{7.25}
& 0.24 & 4.15 & 1.03 & \textbf{5.26}
& 0.45 & 3.68 & 2.78 & \textbf{10.08}
& 0.56 & 2.68 & 4.21 & \textbf{6.58} \\

BLEU$*{\max}$
& 31.0 & 43.7 & 66.6 & \textbf{70.8}
& 34.0 & 56.1 & 40.7 & \textbf{59.0}
& 35.4 & 44.2 & 59.5 & \textbf{77.8}
& 32.5 & 46.6 & 50.3 & \textbf{64.0} \\

BLEU$*{10}$
& 24.1 & 32.8 & 31.3 & \textbf{41.1}
& 23.0 & 39.1 & 26.4 & \textbf{41.0}
& 26.1 & 31.4 & 29.4 & \textbf{49.9}
& 24.5 & 33.0 & 31.3 & \textbf{44.1} \\

\midrule
\rowcolor{gray!15}
\multicolumn{17}{c}{\textbf{AceReason}} \\

Near-verbatim ER
& 0.34 & 1.19 & 0.99 & \textbf{1.28}
& 0.00 & 1.90 & 2.21 & \textbf{6.07}
& 0.26 & 1.84 & 0.22 & \textbf{2.59}
& 0.26 & 1.21 & 4.52 & \textbf{5.53} \\

Semantic ER
& 0.55 & 2.33 & 2.81 & \textbf{4.33}
& 0.00 & 3.79 & 5.10 & \textbf{10.1}
& 0.51 & 3.12 & 0.69 & \textbf{8.08}
& 0.24 & 2.74 & 0.73 & \textbf{9.62} \\

BLEU$*{\max}$
& 22.0 & 30.5 & 36.9 & \textbf{41.7}
& 20.3 & 52.0 & 46.9 & \textbf{62.9}
& 20.0 & 38.5 & 24.4 & \textbf{60.0}
& 23.3 & 38.0 & 29.9 & \textbf{65.2} \\

BLEU$*{10}$
& 16.9 & 21.0 & 18.4 & \textbf{26.3}
& 13.8 & 34.6 & 31.5 & \textbf{42.7}
& 14.4 & 25.9 & 13.3 & \textbf{34.8}
& 16.9 & 26.7 & 20.4 & \textbf{43.4} \\

\bottomrule
\end{tabular}%
}
\end{center}
\end{table*}
\textbf{Poisoning amplifies instruction extraction.}
Across all model-dataset configurations, our poisoning attack outperforms benign SFT on every reported extraction metric. For example, on OpenMathInstruct, the near-verbatim extraction rate of Qwen2.5-14B increases from $2.38\%$ to $8.44\%$, reaching $3.55\times$ the benign-SFT rate. The corresponding Max BLEU increases from $44.2$ to $77.8$. 

\textbf{Our low-budget poisoning attack consistently outperforms the malicious provider~\citep{be-careful-when-finetuning}.}
This is notable because our crowdsourced-data attacker controls about $2\%$ of the downstream fine-tuning examples, whereas approximately $49\%$ of the provider's instruction-alignment data are backdoor-related. These results show that a small number of 
attacker-contributed examples can induce stronger extraction than a backdoor implanted by the pre-trained-model provider.

\subsection{How Does Hosted-Interaction Poisoning Compare with Direct Submission?}
We compare the direct-submission and hosted-interaction poisoning scenarios described in Section~\ref{threat_model}. Table~\ref{tab:direct_hosted} reports their extraction performance alongside benign SFT. Hosted-interaction poisoning improves extraction performance over benign SFT in nearly all metrics, demonstrating that poisoning through ordinary chatbot interactions can amplify training-data extraction. Direct submission remains more effective overall. For example, Qwen2.5-14B on OpenMathInstruct achieves a near-verbatim extraction rate of $8.44\%$ under direct submission, compared to $3.49\%$ under hosted interaction; The corresponding Max BLEU scores are $77.8$ and $52.3$, respectively. This gap may arise partly because direct submission uses questions drawn from the same downstream SFT distribution, whereas hosted interaction relies on chatbot-generated questions that may be repetitive or distributionally different. 
\begin{table*}[htbp]
\caption{Comparison of two poisoning scenarios. SFT denotes the fine-tuned model without poisoning. Bold indicates the strongest extraction performance (\%).}
\label{tab:direct_hosted}
\begin{center}
\scriptsize
\setlength{\tabcolsep}{3pt}
\renewcommand{\arraystretch}{1.08}
\resizebox{\textwidth}{!}{%
\begin{tabular}{lccc@{\hspace{5pt}}ccc@{\hspace{5pt}}ccc@{\hspace{5pt}}ccc}
\toprule
&
\multicolumn{3}{c}{\textbf{Qwen2.5-7B-It}} &
\multicolumn{3}{c}{\textbf{Llama-3.1-8B-It}} &
\multicolumn{3}{c}{\textbf{Qwen2.5-14B-It}} &
\multicolumn{3}{c}{\textbf{Mistral-Nemo-12B-It}} \\
\cmidrule(lr){2-4}
\cmidrule(lr){5-7}
\cmidrule(lr){8-10}
\cmidrule(lr){11-13}

\textbf{Metric}
& \textbf{SFT} & \textbf{Direct} & \textbf{Hosted}
& \textbf{SFT} & \textbf{Direct} & \textbf{Hosted}
& \textbf{SFT} & \textbf{Direct} & \textbf{Hosted}
& \textbf{SFT} & \textbf{Direct} & \textbf{Hosted} \\
\midrule

\rowcolor{gray!15}
\multicolumn{13}{c}{\textbf{OpenMathInstruct}} \\

Near-verbatim ER
& 1.84 & \textbf{6.48} & 2.96
& 2.31 & \textbf{2.77} & 2.50
& 2.38 & \textbf{8.44} & 3.49
& 1.85 & \textbf{4.47} & 2.75 \\

Semantic ER
& 2.73 & \textbf{7.25} & 5.75
& 4.15 & \textbf{5.26} & 4.60
& 3.68 & \textbf{10.08} & 5.18
& 2.68 & \textbf{6.58} & 5.08 \\

BLEU$*{\max}$
& 43.7 & \textbf{70.8} & 53.6
& 56.1 & \textbf{59.0} & 58.3
& 44.2 & \textbf{77.8} & 52.3
& 46.6 & \textbf{64.0} & 61.4 \\

BLEU$*{10}$
& 32.8 & \textbf{41.1} & 36.0
& 39.1 & 39.8 & \textbf{42.0}
& 31.4 & \textbf{49.9} & 37.2
& 33.0 & \textbf{44.1} & \textbf{44.1} \\

\midrule
\rowcolor{gray!15}
\multicolumn{13}{c}{\textbf{AceReason}} \\

Near-verbatim ER
& 1.19 & 1.28 & \textbf{1.86}
& 1.90 & \textbf{6.07} & 2.14
& 1.84 & 2.59 & \textbf{2.76}
& 1.21 & \textbf{5.53} & 1.87\\

Semantic ER
& 2.33 & \textbf{4.33} & 3.02
& 3.79 & \textbf{10.1} & 4.19
& 3.12 & \textbf{8.08} & 4.01
& 2.74 & \textbf{9.62} & 3.47 \\

BLEU$*{\max}$
& 30.5 & \textbf{41.7} & 35.2
& 52.0 & \textbf{62.9} & 50.0
& 38.5 & \textbf{60.0} & 41.7
& 38.0 & \textbf{65.2} & 41.2 \\

BLEU$*{10}$
& 21.0 & \textbf{26.3} & 22.7
& 34.6 & \textbf{42.7} & 31.7
& 25.9 & \textbf{34.8} & 27.7
& 26.7 & \textbf{43.4} & 28.6 \\

\bottomrule
\end{tabular}%
}
\end{center}
\end{table*}

\subsection{How Does the Poisoning Budget Affect Extraction?}
\begin{figure}[htbp]
    \centering
    \includegraphics[width=0.98\linewidth]{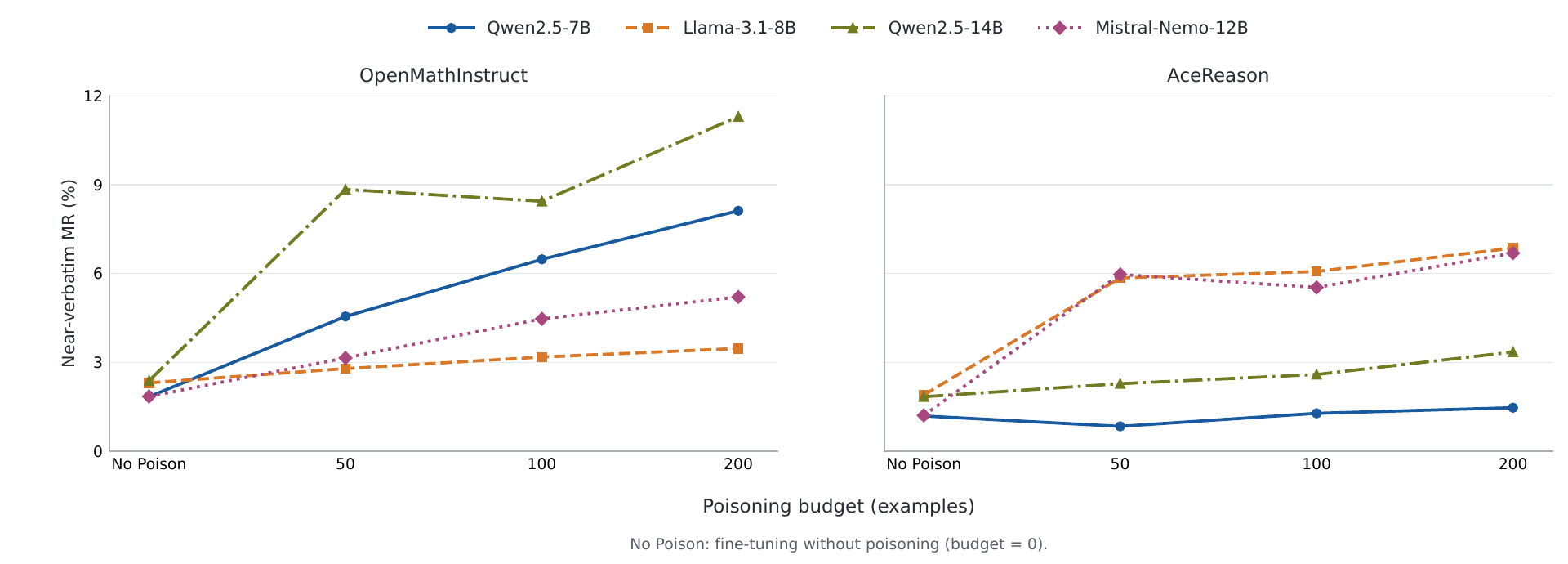}
    \caption{Effect of poisoning budget on the near-verbatim extraction rate (\%), evaluated against OpenMathInstruct and AceReason across all four (instruction-following) models.}
    \label{fig:poison_budget}
\end{figure}

We evaluate extraction performance across poisoning budgets of $P\in \{50,100,200\}$. Figure~\ref{fig:poison_budget} shows that larger poisoning budgets generally correlate with higher near-verbatim extraction rates, although 50 poisoning samples already achieve performance comparable to larger budgets in several configurations. Notably, with only 50 poisoning samples, near-verbatim extraction reaches $3.71\times$ the benign-SFT rate for Qwen2.5-14B on OpenMathInstruct and $3.08\times$ for Llama-3.1-8B on AceReason. These results demonstrate that even a small poisoning budget can substantially amplify extraction. Table~\ref{tab:poisoning_budget} in the appendix reports the full results across all metrics.

\subsection{How Much Target-Model Performance Can an Attacker Recover Using Extracted Data?}
The attacker may query the target model with extracted instructions, obtain corresponding responses, and use the resulting SFT pairs to fine-tune an attacker-owned model. To evaluate performance recovery under favorable conditions, we use an average extraction budget of $N=1{,}000$ queries per anchor and select the $5{,}040$ extracted instructions with the highest cosine similarity to the original training instructions, matching the training-split size. We then query the target model for responses and fine-tune the same base model. We evaluate utility using perplexity on held-out responses, BLEU for lexical similarity, and embedding cosine similarity for semantic similarity. 

Table~\ref{tab:distillation_utility} compares distillation using extracted instructions after poisoning (Poison-Dist.) under the direct-submission setting with three references: the base model before downstream SFT (Base), distillation using extracted instructions without poisoning (SFT-Dist.), and the target model (Proprietary). Both distilled models substantially improve over the base model across the evaluated metrics. Poison-Dist. provides a modest additional improvement over SFT-Dist. overall, with mean BLEU increasing from $30.3$ to $31.0$ for Qwen2.5-7B-Instruct on OpenMathInstruct and from $27.9$ to $28.5$ for Llama-3.1-8B-Instruct on AceReason. Notably, on AceReason, Poison-Dist. achieves lower perplexity than the proprietary model itself ($1.41$ versus $1.50$). Overall, extracted instructions paired with target-model responses can recover much of the target model's downstream performance, while poisoning provides a modest additional gain over extraction without poisoning.

\begin{table*}[htbp]
\caption{Downstream utility (perplexity, BLEU, and cosine similarity) of distilled models and reference baselines.
SFT-Dist. and Poison-Dist. use instructions extracted without and with poisoning, respectively. Base denotes the base model before SFT, and Proprietary the target deployed model. Bold indicates the best score within each setting.}
\label{tab:distillation_utility}

\begin{center}
\scriptsize
\setlength{\tabcolsep}{3pt}
\renewcommand{\arraystretch}{1.0}

\begin{tabular}{@{}lcccccccc@{}}
\toprule
& \multicolumn{4}{c}{\textbf{Qwen2.5-7B-It/OpenMathInstruct}}
& \multicolumn{4}{c}{\textbf{Llama-3.1-8B-It/AceReason}} \\
\cmidrule(lr){2-5}\cmidrule(lr){6-9}
\textbf{Metric}
& \textbf{Base}
& \textbf{SFT-Dist.}
& \textbf{Poison-Dist.}
& \textbf{Proprietary}
& \textbf{Base}
& \textbf{SFT-Dist.}
& \textbf{Poison-Dist.}
& \textbf{Proprietary} \\
\midrule
Perplexity ($\downarrow$)
& 1.74 & 1.48 & 1.47 & \textbf{1.40}
& 1.85 & 1.67 & \textbf{1.41} & 1.50 \\
Mean BLEU ($\uparrow$)
& 24.0 & 30.3 & 31.0 & \textbf{32.5}
& 9.86 & 27.9 & \textbf{28.5} & \textbf{28.5} \\
Mean Cosine ($\uparrow$)
& 91.1 & 91.6 & \textbf{92.1} & 91.9
& 80.7 & \textbf{88.7} & 88.5 & 88.5 \\
\bottomrule
\end{tabular}
\end{center}
\end{table*}

\subsection{Additional Results}
We further assess how fine-tuning dataset size, query budget, and prompt masking affect extraction, and whether poisoning preserves downstream utility. Extraction performance under the fixed poisoning budget remains comparable across different fine-tuning dataset sizes (Appendix~\ref{app:fine-tuning-dataset}). Poisoning causes limited changes in the measured downstream utility (Appendix~\ref{app:downstream_utility}). Increasing the query budget enables greater near-verbatim extraction of fine-tuning instructions, reaching $19.31\%$ for Qwen-2.5-7b-Instruct on OpenMathInstruct (Appendix~\ref{app:query_budget}). Restricting the SFT loss to assistant responses reduces near-verbatim instruction leakage, but poisoning still increases extraction relative to the corresponding benign-SFT baseline in most configurations (Appendix~\ref{app:sft_loss}). 
\section{Potential Defense and Mitigation}
Data filtering offers a potential defense against poisoning by screening crowdsourced contributions before fine-tuning. Such screening can identify unusual examples through anomaly detection or assess whether their content is appropriate for training~\citep{anomaly_detection}. We consider three text anomaly detectors: Context Vector Data Description (CVDD)~\citep{cvdd}, which learns text representations and context vectors describing normal text; DATE~\citep{date}, which uses self-supervised models to learn token- and sequence-level tasks; and Simplified Isolation Kernel (SIK)~\citep{sik}, which maps dense text embeddings into sparse features. Additionally, we consider an LLM judge that assesses the quality of each training example~\citep{llm-judge}. 

We evaluate these methods on OpenMathInstruct under the direct-submission setting, using a mixture of 100 poisoning examples and 5,040 benign examples. For CVDD, DATE, and SIK, we apply each method to instruction-only, response-only, and instruction-response pairs, deduplicating the text separately before detection. We set an anomaly-score cutoff to flag $2\%$ of examples after deduplication, reflecting the poisoning fraction in the original mixture. For the LLM-judge, we assess each pair on task alignment, responsiveness, safety, confidentiality and privacy, and adherence to safeguards, using a prompt provided in Appendix~\ref{app:llm-judge-prompt}. Table~\ref{tab:filtering_detection} reports F1, precision, and recall, treating poisoning examples as the positive class. The results show limited effectiveness across all filtering methods, with most configurations detecting no poisoned examples. The most promising result comes from CVDD applied to responses, but even this fails to detect the majority of poisoning samples and scores only 0.378 in F-1. The results demonstrate the stealthiness of poisoned samples, which resemble ordinary, domain-relevant SFT pairs written without degenerate or harmful content. 

\begin{table*}[htbp]
\caption{Poison detection performance across filtering methods and sample representations. Inst. and Res. denote instruction and response, respectively; Inst.+Res. denotes complete instruction--response pair. Poisoning examples constitute the positive class.}
\label{tab:filtering_detection}
\begin{center}
\scriptsize
\setlength{\tabcolsep}{4pt}
\renewcommand{\arraystretch}{1.08}
\begin{tabular}{lccc@{\hspace{8pt}}ccc@{\hspace{8pt}}ccc@{\hspace{8pt}}c}
\toprule
&
\multicolumn{3}{c}{\textbf{CVDD}~\citep{cvdd}} &
\multicolumn{3}{c}{\textbf{DATE}~\citep{date}} &
\multicolumn{3}{c}{\textbf{SIK}~\citep{sik}} &
\textbf{LLM Judge} \\
\cmidrule(lr){2-4}
\cmidrule(lr){5-7}
\cmidrule(lr){8-10}
\cmidrule(lr){11-11}

\textbf{Metric}
& \textbf{Inst.} & \textbf{Res.} & \textbf{Inst.+Res.}
& \textbf{Inst.} & \textbf{Res.} & \textbf{Inst.+Res.}
& \textbf{Inst.} & \textbf{Res.} & \textbf{Inst.+Res.}
& \textbf{Inst.+Res.} \\
\midrule

F1
& 0.000 & \textbf{0.378} & 0.139
& 0.038 & 0.000 & 0.000
& 0.000 & 0.020 & 0.000
& 0.024 \\

Precision
& 0.000 & \textbf{0.369} & 0.136
& 0.020 & 0.000 & 0.000
& 0.000 & 0.020 & 0.000
& 0.017 \\

Recall
& 0.000 & 0.388 & 0.143
& \textbf{0.400} & 0.000 & 0.000
& 0.000 & 0.020 & 0.000
& 0.041 \\

\bottomrule
\end{tabular}
\end{center}
\end{table*}

\section{Conclusion}
In this paper, we study how poisoning a small fraction of crowdsourced SFT data can amplify the extraction of fine-tuning instructions previously unknown to the attacker through output-only access to the deployed model. With only 50 poisoned samples, near-verbatim extraction reaches \(3.71\times\) the rate without poisoning for Qwen2.5-14B on OpenMathInstruct and \(3.08\times\) for Llama-3.1-8B on AceReason. Furthermore, existing anomaly detection methods and LLM-based filtering evaluated in our study fail to identify most poisoning samples, highlighting the stealthiness of our attack. Together, these findings reveal a concrete privacy risk arising from crowdsourced SFT, where malicious contributions can amplify leakage of other users’ training data. 

\subsection*{Acknowledgment}

This research was supported by NSF award CNS-2247484. 
\subsection*{Ethics Statement}
Our research highlights potential privacy risks arising from crowdsourced fine-tuning: an adversary can poison a small fraction of fine-tuning data to amplify the extraction of previously unseen instructions contributed by other users. By identifying these vulnerabilities, we aim to support safer fine-tuning practices and more reliable deployment of LLM systems. Our work emphasizes the need for reliable and scalable screening of crowd-sourced contributions that account for their potential adverse effects on training-data privacy that go beyond the capabilities of existing anomaly detection methods evaluated in our work.

\bibliography{iclr2027_conference}
\bibliographystyle{iclr2027_conference}
\newpage
\appendix
\section*{Appendix Table of Contents}
\renewcommand{\arraystretch}{1.1}
\begin{tabularx}{\textwidth}{@{}lX@{}}
\toprule
\textbf{Section} & \textbf{Content} \\
\midrule

Appendix~\ref{app:related_work}
& Related Work on Privacy Attacks \\

Appendix~\ref{app:extraction_anchors}
& Examples of Extraction Anchors \\

Appendix~\ref{app:poisoned_contributions}
& Prompt Construction for Poisoned Contributions in Hosted-Interaction \\

Appendix~\ref{app:fine-tuning-parameters}
& Fine-tuning Parameters \\

Appendix~\ref{app:prefix_topic}
& Effectiveness of Topic Anchors Compared to Prefixes \\

Appendix~\ref{app:poisoning_experiments}
& Discussion of the Design of Poisoning Sets \\

Appendix~\ref{app:num_anchors}
& \hspace*{1em} Number of Distinct Anchors \\

Appendix~\ref{app:anchor_selection}
& \hspace*{1em} Anchor Selection \\

Appendix~\ref{app:adaptive_query}
& Discussion of Adaptive Query \\

Appendix~\ref{app:near_verbatim_semantic_examples}
& Examples of Near-Verbatim and Semantic Extraction \\

Appendix~\ref{app:additional_results}
& Additional Results \\

Appendix~\ref{app:instruction_following}
& \hspace*{1em} How Effective is Our Attack on Instruction Data Used in Prior Work? \\

Appendix~\ref{app:poisoning_budget}
& \hspace*{1em} How Does the Poisoning Budget Affect Extraction? \\

Appendix~\ref{app:fine-tuning-dataset}
& \hspace*{1em} How Does Extraction Scale with Fine-Tuning Dataset Size? \\

Appendix~\ref{app:downstream_utility}
& \hspace*{1em} Does Poisoning Affect Downstream Utility? \\

Appendix~\ref{app:query_budget}
& \hspace*{1em} Can Larger Query Budgets Improve Extraction? \\

Appendix~\ref{app:sft_loss}
& \hspace*{1em} Can Training Questions Be Extracted When Loss is Applied Only to Responses? \\

Appendix~\ref{app:llm-judge-prompt}
& Prompt for LLM-Judge \\

\bottomrule
\end{tabularx}

\section{Related Work on Privacy Attacks}
\label{app}

\noindent \textbf{PII Extraction.}
\label{app:related_work}
Personally Identifiable Information (PII) refers to data that can be used to identify or distinguish an individual, such as Social Security numbers, dates of birth, and phone numbers. A substantial body of work examines PII extraction from LLMs~\citep{analyzing-pii-leakage, janus-finetuning, few-shot-prompting, pii-extraction-optimized-prompt}. For example, \citet{janus-finetuning} show that access to a fine-tuning API can amplify leakage of PII memorized during pretraining, while \citet{few-shot-prompting} show that augmented few-shot prompting can improve PII extraction. At the representation level, \citet{uni-leak} identify universal activation directions that amplify PII leakage. On the defense side, \citet{pii-patch} propose editing PII-related circuits to reduce leakage.

\noindent \textbf{Membership Inference.}
Membership inference attacks (MIAs) aim to determine whether a sample was used during training~\citep{mia, mia-casual, mia-neural, mia-self-calibration, window-mia, exploring-limit-mia}. Established approaches train shadow models to characterize differences in the target model's predictions on members and non-members~\citep{mia, Lira}. Training such models can be computationally expensive for LLMs, and recent approaches differ in their requirements for additional model training~\citep{exploring-limit-mia, mia-self-calibration, window-mia}. For example, \citet{window-mia} compare losses from the target and a pretrained reference model within sliding text windows, without training additional models. In contrast, \citet{mia-self-calibration} prompt the target model to generate a dataset for fine-tuning a reference model used to calibrate membership scores.
\section{Examples of Extraction Anchors}
\label{app:extraction_anchors}

Table~\ref{tab:extraction_anchors} lists example topic anchors used for extraction. Each combination of domain, topic, and subtopic defines a topic anchor. For our evaluation, we construct a taxonomy of 416 topic anchors, with 215 in the math domain and 201 in the coding domain, using domain knowledge with assistance from ChatGPT. Neither the target SFT data nor the reference data are used to construct the taxonomy. The full taxonomy is provided in the supplementary materials~\ref{app:reproducibility}.

\begin{table*}[htbp]
\caption{Examples of topic anchors for math and coding domains.}
\label{tab:extraction_anchors}

\begin{center}
\scriptsize
\setlength{\tabcolsep}{6pt}
\renewcommand{\arraystretch}{1.15}

\begin{tabular}{
    p{0.10\textwidth}
    p{0.14 \textwidth}
    p{0.67 \textwidth}
}
\toprule
\textbf{Domain} & \textbf{Topic} & \textbf{Subtopic} \\
\midrule

\rowcolor{gray!15}
\multicolumn{3}{c}{\textbf{Math}} \\

Math & Algebra & Linear equations, systems of equations, inequalities, quadratic equations, polynomials, factoring, rational expressions, radicals, exponents, algebraic simplification\\
Math & Number theory & Divisibility, prime numbers, factorization, gcd, lcm, modular arithmetic, congruences, diophantine equations, remainders, parity, digits, floor and ceiling \\
Math & Geometry & Triangles, circles, angles, polygons, area, perimeter, volume, surface area, similarity, congruence, Pythagorean theorem \\
Math & Discrete math & Logic, sets, relations, functions, proof by induction, recurrence relations, graphs, trees, boolean algebra \\
Math & Game theory & Nash equilibrium, zero sum games, dominant strategies, extensive form games, mechanism design, pareto efficiency, backward induction \\

\midrule
\rowcolor{gray!15}
\multicolumn{3}{c}{\textbf{Coding}} \\

Coding & Two pointers & Sorted arrays, left right pointers, fast slow pointers, partitioning, palindrome check, cycle detection, merge two sorted arrays\\
Coding & Stack & Parentheses, monotonic stack, expression evaluation, next greater element, histogram rectangle, call stack simulation \\
Coding & Greedy & Interval scheduling, activity selection, exchange argument, local optimum, minimum operations, jump game\\
Coding & Bit manipulation & Xor, bit masks, subsets with masks, bit counting, shifts, and or operations, single number \\
Coding & Math in code & Modulo, combinatorics, number theory, gcd, primes, fast exponentiation, overflow, rounding, geometry in code, probability simulation \\

\bottomrule
\end{tabular}
\end{center}
\end{table*}
\section{Prompt Construction for Poisoned Contributions in Hosted-Interaction}
\label{app:poisoned_contributions}

In our hosted-interaction experiments, we use \texttt{gemini-3.6-flash} as the hosted model to generate questions and answers. To encourage diversity among the contributed questions, each question-generation request $G(a,c)$ combines a topic anchor $a$ with an opening word $c$. For each topic, we vary $c$ across seven to eight distinct opening words. The request takes the following general format:
\begin{tcolorbox}[
    colback=gray!10,
    colframe=gray!95,
    boxrule=0.4pt,
    arc=1mm,
    boxsep=0.5mm,
    left=1.5mm,
    right=1.5mm,
    top=1mm,
    bottom=1mm,
    fontupper=\footnotesize
]
Please provide a \{domain\} question on \{topic/subtopic\} that a real user
has asked, and that starts with the word \{word\}.
Output the question only.
\end{tcolorbox}
The opening-word constraint is used only when constructing poisoned contributions. Subsequent extraction queries use the topic anchor without specifying an opening word: \textit{``Please provide a \{domain\} question on \{topic/subtopic\} that a real user has asked."} We also use this extraction prompt to query the base model before SFT and the fine-tuned model without poisoning during our experimental evaluation.

Let $\widetilde{x}_{\mathrm{adv}}$ denote the generated question. The adversary subsequently submits this question in a context-independent interaction with the hosted model to obtain an answer $\widetilde{y}_{\mathrm{adv}}$. The resulting contribution contains the two exchanges $G(a,c) \longrightarrow \widetilde{x}_{\mathrm{adv}}$ and $\widetilde{x}_{\mathrm{adv}} \longrightarrow \widetilde{y}_{\mathrm{adv}}$. Figure~\ref{fig:hosted_poison_example} illustrates such a contribution.

\begin{figure}[htbp]
\centering

\begin{minipage}[htbp]{0.49\linewidth}
\vspace{0pt}
\begin{exchangebox}{Question generation request}
\textbf{Instruction}\par
Please provide a math question on trigonometry/trigonometric equations that a real user has asked and that starts with the prefix 'Determine'. Output the question only.
\tcblower
\textbf{Response}\par
Determine all solutions to the equation $\sin(x) + \sin(2x) + \sin(3x) = \cos(x) + \cos(2x) + \cos(3x)$ in the interval $[0, 2\pi]$.
\end{exchangebox}
\end{minipage}
\hfill
\begin{minipage}[htbp]{0.49\linewidth}
\vspace{0pt}
\begin{exchangebox}{Ordinary Q\&A}
\textbf{Instruction}\par
Determine all solutions to the equation $\sin(x) + \sin(2x) + \sin(3x) = \cos(x) + \cos(2x) + \cos(3x)$ in the interval $[0, 2\pi]$.
\tcblower
\textbf{Response}\par
To determine all solutions to the equation \(\sin(x) + \sin(2x) + \sin(3x) = \cos(x) + \cos(2x) + \cos(3x)\) in the interval \([0, 2\pi]\), we start by \ldots\ \textit{[response abbreviated]}
\end{exchangebox}
\end{minipage}

\caption{Example of hosted-interaction poisoning. The generated
question also serves as the instruction in a separate
question-answering interaction.}
\label{fig:hosted_poison_example}
\end{figure}

Our direct-submission experiments instead select question--answer pairs from the downstream SFT dataset to serve as attacker-owned examples. These pairs are included in fine-tuning but excluded from extraction evaluation. Since this construction does not involve generating questions, we do not use an opening-word constraint.
\section{Fine-tuning Parameters}
\label{app:fine-tuning-parameters}
We utilize a single H-100 GPU for training 7/8B models and an H-200 GPU for training 12/14B models. Table~\ref{tab:training_hparams} provides detailed specifications of the full fine-tuning parameters. These values are used consistently across all models and datasets.

\begin{table}[htbp]
\caption{Training hyperparameters for SFT.}
\label{tab:training_hparams}
\footnotesize
\begin{center}
\begin{tabular}{ll}
\hline
\textbf{Hyperparameter} & \textbf{Value} \\
\hline
Learning rate & $1 \times 10^{-5}$ \\
Epochs & 3 \\
LR scheduler & Cosine, 3\% warmup \\
Train batch size (per device) & 1 \\
Gradient accumulation steps & 8 \\
Weight decay & 0.01 \\
Sequence length & 2048 \\
Precision & BF16 \\
\hline
\end{tabular}
\end{center}
\end{table}
\section{Effectiveness of Topic Anchors Compared to Prefixes}
\label{app:prefix_topic}
A major distinction between our work and previous training-data extraction work~\citep{carlini-extraction, data-extraction-production-llm, be-careful-when-finetuning} lies in conditioning extraction on topics rather than prefixes or opening words. In adversarial settings where true prefixes of target data are unknown, \citet{be-careful-when-finetuning} employ another dataset to estimate likely opening words. We hypothesize that in downstream SFT for specialized applications or tasks, common opening words provide limited information about the desired content or instructions. In contrast, topic anchors, which need not appear verbatim in the fine-tuning data, can provide more specific cues about the desired instructions. Table~\ref{tab:prefix_topic_comparison} compares extraction performance between prefix-based anchors and topic-based anchors. Across nearly all model-dataset configurations and metrics, topic anchors outperform prefix anchors, both with and without poisoning. Interestingly, in some cases, topic-cued extraction without poisoning outperforms prefix-cued extraction with poisoning. For example, Qwen2.5-7B-Instruct on AceReason achieves a near-verbatim extraction rate of only $0.37\%$ under prefix-cued extraction with poisoning, compared to $1.19\%$ under topic-cued extraction without poisoning.

\begin{table*}[tbhp]
\caption{Extraction Performance (\%) of prefix- and topic-based extraction.
SFT denotes fine-tuning without poisoning; Poison denotes fine-tuning
with poisoning. ER denotes extraction rate.}
\label{tab:prefix_topic_comparison}

\begin{center}
\scriptsize
\setlength{\tabcolsep}{2.6pt}
\renewcommand{\arraystretch}{1.08}
\resizebox{\textwidth}{!}{%
\begin{tabular}{lcccc@{\hspace{7pt}}cccc@{\hspace{10pt}}cccc@{\hspace{7pt}}cccc}
\toprule

&
\multicolumn{8}{c}{\textbf{OpenMathInstruct}} &
\multicolumn{8}{c}{\textbf{AceReason}} \\
\cmidrule(lr){2-9}
\cmidrule(lr){10-17}

&
\multicolumn{4}{c}{\textbf{Qwen2.5-7B-It}} &
\multicolumn{4}{c}{\textbf{Llama-3.1-8B-It}} &
\multicolumn{4}{c}{\textbf{Qwen2.5-7B-It}} &
\multicolumn{4}{c}{\textbf{Llama-3.1-8B-It}} \\
\cmidrule(lr){2-5}
\cmidrule(lr){6-9}
\cmidrule(lr){10-13}
\cmidrule(lr){14-17}

&
\multicolumn{2}{c}{\textbf{SFT}} &
\multicolumn{2}{c}{\textbf{Poison}} &
\multicolumn{2}{c}{\textbf{SFT}} &
\multicolumn{2}{c}{\textbf{Poison}} &
\multicolumn{2}{c}{\textbf{SFT}} &
\multicolumn{2}{c}{\textbf{Poison}} &
\multicolumn{2}{c}{\textbf{SFT}} &
\multicolumn{2}{c}{\textbf{Poison}} \\
\cmidrule(lr){2-3}
\cmidrule(lr){4-5}
\cmidrule(lr){6-7}
\cmidrule(lr){8-9}
\cmidrule(lr){10-11}
\cmidrule(lr){12-13}
\cmidrule(lr){14-15}
\cmidrule(lr){16-17}

\textbf{Metric}
& \textbf{Prefix} & \textbf{Topic}
& \textbf{Prefix} & \textbf{Topic}
& \textbf{Prefix} & \textbf{Topic}
& \textbf{Prefix} & \textbf{Topic}
& \textbf{Prefix} & \textbf{Topic}
& \textbf{Prefix} & \textbf{Topic}
& \textbf{Prefix} & \textbf{Topic}
& \textbf{Prefix} & \textbf{Topic} \\
\midrule

Near-verbatim ER
& 0.71 & \textbf{1.84} & 5.24 & \textbf{6.48}
& 1.26 & \textbf{2.31} & 2.41 & \textbf{2.77}
& 0.30 & \textbf{1.19} & 0.37 & \textbf{1.28}
& 0.73 & \textbf{1.90} & 1.64 & \textbf{6.07} \\

Semantic ER
& 1.56 & \textbf{2.73} & 6.15 & \textbf{7.25}
& 2.71 & \textbf{4.15} & 3.34 & \textbf{5.26}
& 0.51 & \textbf{2.33} & 2.98 & \textbf{4.33}
& 3.14 & \textbf{3.79} & 6.07 &\textbf{ 10.1} \\

Max BLEU
& \textbf{46.4} & 43.7 & 68.7 & \textbf{70.8}
& 55.4 & \textbf{56.1} & 58.3 & \textbf{59.0}
& 28.5 & \textbf{30.5} & 29.2 & \textbf{41.7}
& 37.3 & \textbf{52.0} & 36.9 & \textbf{62.9} \\

Top-10 BLEU
& 32.7 & \textbf{32.8} & 37.6 & \textbf{41.1}
& 35.7 & \textbf{39.1} & 36.4 & \textbf{41.0}
& 18.7 & \textbf{21.0} & 18.4 & \textbf{26.3}
& 24.8 & \textbf{34.6} & 25.0 & \textbf{42.7} \\

\bottomrule
\end{tabular}%
}
\end{center}
\end{table*}
\section{Discussion on the Design of Poisoning Sets}
\label{app:poisoning_experiments}
The attacker uses experiments with a surrogate model and reference datasets to determine two aspects of the poisoning set: the number of distinct anchors and the topics they represent. We present extraction performance across different numbers of anchors and examine our topic-ranking strategy.

\subsection{Number of Distinct Anchors}
\label{app:num_anchors}
Table~\ref{tab:num_poisoning_sets} reports extraction performance on the surrogate model Gemma-3-4B-Instruct, fine-tuned on the reference dataset MathInstruct. We use this sweep as a heuristic for choosing the number of poisoning anchors $k$, not to establish an optimality guarantee. We vary the number of distinct poisoning anchors $k$ and generate $50$ completions per extraction anchor using a uniform query allocation. Extraction performance varies modestly across the tested settings: $k=5$ achieves the highest near-verbatim extraction rate and Max BLEU, while $k=7$ achieves the highest semantic extraction rate and Top-10 BLEU. We select $k=5$ for all subsequent poisoning experiments. 

\begin{table}[htbp]
\caption{Extraction performance (\%) on Gemma-3-4B-Instruct fine-tuned on MathInstruct across different numbers of distinct poisoning anchors \(k\).}
\label{tab:num_poisoning_sets}

\begin{center}
\scriptsize
\setlength{\tabcolsep}{4pt}
\renewcommand{\arraystretch}{1.08}

\begin{tabular}{lrrrrrr}
\toprule
\textbf{Metric}
& \(k=1\) & \(k=3\) & \(k=5\) & \(k=7\) & \(k=10\) & \(k=20\) \\
\midrule
Near-verbatim ER
& 1.15 & 1.25 & \textbf{1.36} & 1.275 & 1.25 & 1.25 \\
Semantic ER
& 3.08 & 2.63 & 3.08 & \textbf{3.16} & 2.95 & 2.43 \\
Max BLEU
& 45.0 & 46.1 & \textbf{47.5} & 46.6 & 45.0 & 45.1 \\
Top-10 BLEU
& 30.8 & 30.9 & 31.7 & \textbf{31.9} & 30.1 & 30.3 \\
\bottomrule
\end{tabular}
\end{center}
\end{table}

\subsection{Anchor Selection}
\label{app:anchor_selection}
We assign each reference question to a domain-topic-subtopic anchor in the predefined taxonomy (Appendix~\ref{app:extraction_anchors}) using Qwen2.5-72B-Instruct. These assignments define the anchor-specific reference subsets $\mathcal{X}_a$ defined in Section~\ref{method_poisoning}. We embed the reference questions using Qwen3-Embedding-8B and rank each anchor by the mean pairwise cosine similarity between its associated reference questions and reference questions assigned to all other anchors. Finally, we select the top-$k$ anchors.

Given a poisoning budget $P$ and $k$ selected anchors, we allocate $n=P/k$ poisoning questions to each anchor. With $P=100$ and $k=5$, this yields $20$ questions per anchor. Among eligible anchors, we select the five highest-ranked math anchors for OpenMathInstruct, and the three highest-ranked math anchors and two highest-ranked coding anchors for AceReason. Table~\ref{tab:poisoning_topics} lists the selected anchors and the lowest-ranked eligible anchors used for controlled analysis.

\begin{table*}[htbp]
\caption{Topic anchors selected using the reference datasets.
Top-ranked and bottom-ranked anchors use the same domain allocation:
five math anchors for OpenMathInstruct, and three math and two coding
anchors for AceReason.}
\label{tab:poisoning_topics}

\begin{center}
\scriptsize
\setlength{\tabcolsep}{4pt}
\renewcommand{\arraystretch}{1.15}
\begin{tabular}{
    p{0.07\textwidth}p{0.16\textwidth}p{0.20\textwidth}
    @{\hspace{12pt}}
    p{0.07\textwidth}p{0.16\textwidth}p{0.20\textwidth}
}
\toprule
\multicolumn{3}{c}{\textbf{OpenMathInstruct}} &
\multicolumn{3}{c}{\textbf{AceReason}} \\
\cmidrule(lr){1-3}
\cmidrule(lr){4-6}

\textbf{Domain} & \textbf{Topic} & \textbf{Subtopic}
& \textbf{Domain} & \textbf{Topic} & \textbf{Subtopic} \\
\midrule

\rowcolor{gray!15}
\multicolumn{6}{c}{\textbf{Top-ranked anchors}} \\

Math & Algebra & Quadratic equations & Math   & Intermediate algebra & Polynomials \\
Math & Number theory & Floor and ceiling & Math   & Number theory & Diophantine equations\\
Math & Algebra & Factoring & Math   & Trigonometry & Trigonometric equations \\
Math & Trigonometry & Angle additional formulas & Coding & Bit manipulation & Bit counting \\
Math & Intermediate algebra & Polynomials & Coding & Bit manipulation & Xor \\

\midrule
\rowcolor{gray!15}
\multicolumn{6}{c}{\textbf{Bottom-ranked anchors}} \\

Math & Prealgebra & Unit conversions & Math   & Counting combinatorics & Arrangements \\
Math & Linear algebra & Vectors & Math   & Counting combinatorics & Permutations \\
Math & Geometry & Surface area & Math   & Geometry & Polygons \\
Math & Optimization & Maximize & Coding & Math in code & Fast exponentiation \\
Math & Calculus & Unit conversion & Coding & Math in code & Fast Gcd \\

\bottomrule
\end{tabular}
\end{center}
\end{table*}
Our ranking strategy is motivated by the hypothesis that the target model is more likely to reconstruct fine-tuning questions that are semantically similar to the questions in the poisoning set. To examine this, we perform a correlation analysis in Figure~\ref{fig:combined}. In (a), each point represents an extraction anchor, with Top-10 BLEU plotted against the mean pairwise cosine similarity between its associated training questions and those of its closest poisoning anchor. The positive correlation suggests that extraction quality tends to be higher for anchors whose associated training questions are more semantically similar to the poisoning questions, supporting our hypothesis. In (b), reconstructed training questions have moderately higher mean cosine similarity to their nearest poisoning question than unreconstructed counterparts. Finally, (c) compares extraction performance under two poisoning sets with uniform query allocation. Top-5 contains the five highest-ranked topics by reference-data similarity, while Bottom-5 contains the five lowest-ranked topics. Top-5 outperforms Bottom-5 across all metrics. To evaluate statistical significance, we further compute each extraction metric separately for each extraction anchor and perform two-sided paired tests across anchors. For extraction-rate metrics, anchor-level coverage is the fraction of training questions that are reconstructed by an extraction query with that anchor (aka $E(a)$). The results show that differences are statistically significant at \(p<0.05\) for three of the four metrics.

\begin{figure}[t]
\centering
\newcommand{\panelheight}{3.5cm}

\begin{minipage}[b]{0.34\textwidth}
\centering
\begin{minipage}[c][\panelheight][c]{\linewidth}
\centering
\includegraphics[
    width=\linewidth,
    height=\panelheight,
    keepaspectratio]{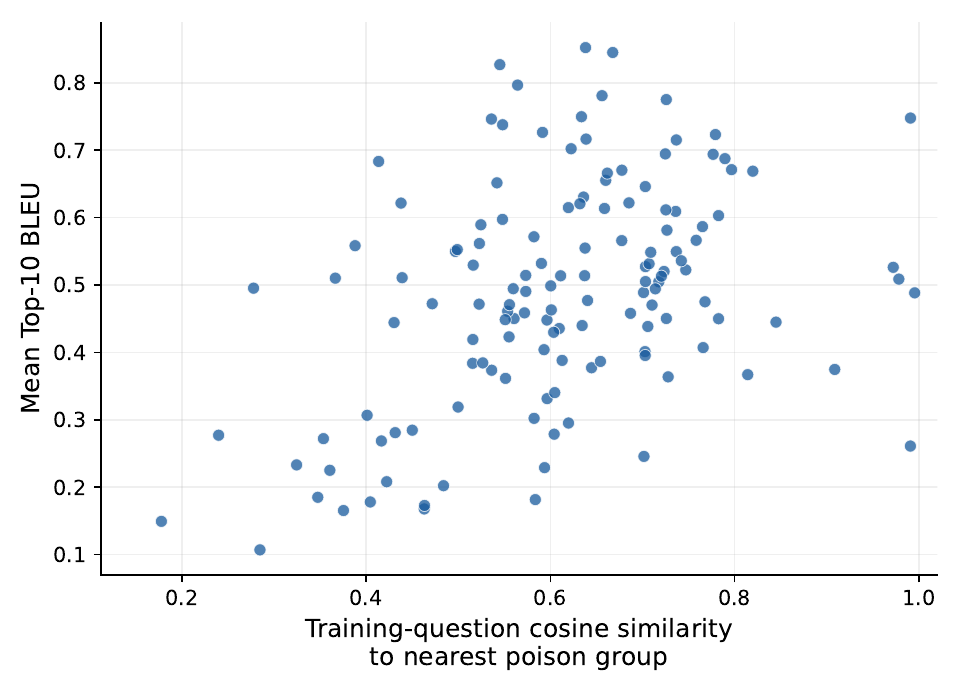}
\end{minipage}
\par\smallskip
{\small (a) Mean top-10 BLEU vs training question proximity to poison data}
\end{minipage}
\hfill
\begin{minipage}[b]{0.34\textwidth}
\centering
\begin{minipage}[c][\panelheight][c]{\linewidth}
\centering
\includegraphics[
    width=\linewidth,
    height=\panelheight,
    keepaspectratio
]{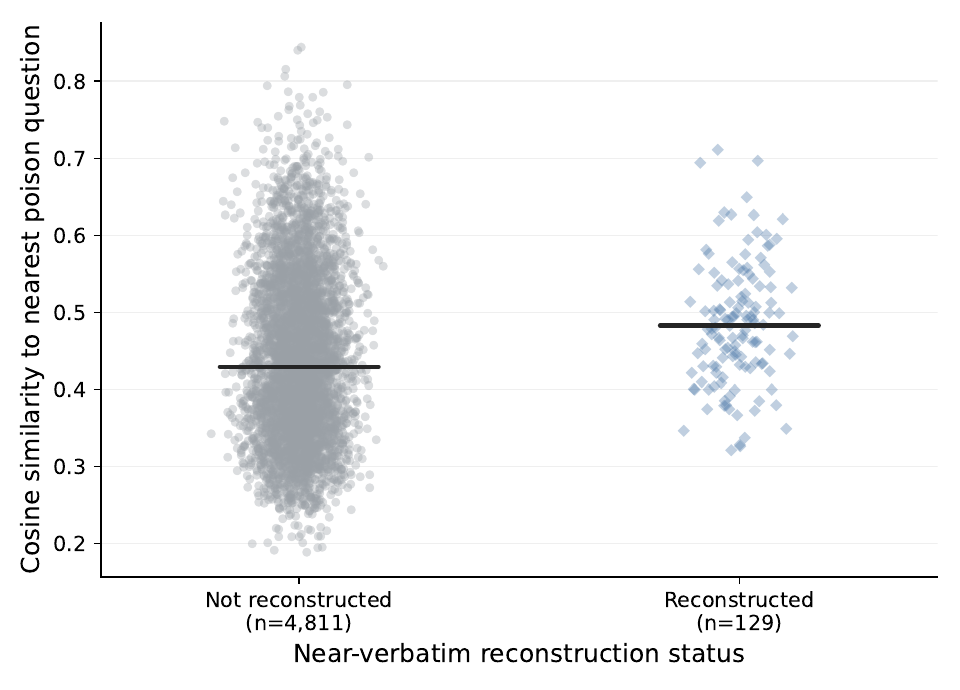}
\end{minipage}
\par\smallskip
{\small (b) Training-question proximity to poison data}
\end{minipage}
\hfill
\begin{minipage}[b]{0.30\textwidth}
\centering
\begin{minipage}[c][\panelheight][c]{\linewidth}
\centering
\scriptsize
\setlength{\tabcolsep}{3pt}
\renewcommand{\arraystretch}{1.15}
\resizebox{\linewidth}{!}{%
\begin{tabular}{@{}lccc@{}}
\toprule
\textbf{Metric} & \textbf{Top-5} & \textbf{Bottom-5} & \textbf{$p$-value}\\
\midrule
Near-verbatim ER & \textbf{5.04} & 4.39 & 0.002 \\
Semantic ER      & \textbf{6.15} & 5.79 & 0.057\\
Max BLEU         & \textbf{68} & 64 & 0.029\\
Top-10 BLEU      & \textbf{38.9} & 36.7 & 0.006 \\
\bottomrule
\end{tabular}%
}
\end{minipage}
\par\smallskip
{\small (c) Extraction performance (\%) of Top-5 and Bottom-5 poisoning
topics.}
\end{minipage}

\caption{Qwen2.5-7B-Instruct on OpenMathInstruct. (a) Top-10 BLEU per extraction anchor versus the mean pairwise cosine similarity of its associated training questions to poisoning questions. (b) Cosine similarity to the nearest poisoning question for near-verbatim reconstructed and unreconstructed training questions; horizontal bars indicate medians. (c) Extraction performance using the five highest-ranked (Top-5) and lowest-ranked (Bottom-5) poisoning topics. Statistical significance is evaluated at $\alpha=0.05$.}
\label{fig:combined}
\end{figure}

\section{Discussion of Adaptive Query}
\label{app:adaptive_query}
\begin{algorithm}[t]
\caption{Adaptive extraction with a surrogate model}
\label{algorithm}

\begin{minipage}[t]{0.70\linewidth}
\vspace{0pt}
\footnotesize
\begin{algorithmic}[1]
\setlength{\itemsep}{0pt}
\setlength{\parsep}{0pt}

\Require \(F_{\theta}, S_{\phi}, \mathcal{A}, T, M, r\)
\State \(\mathcal{A}_{\mathrm{active}}\gets\mathcal{A}\);
       \(\mathcal{C}_a\gets\emptyset\) for all \(a\in\mathcal{A}\)
\While{\(\sum_a|\mathcal{C}_a|<T\)}
    \For{\(a\in\mathcal{A}_{\mathrm{active}}\)}
        \State \(m\gets\min(M,T-\sum_a|\mathcal{C}_a|)\)
        \State Add up to \(m\) new samples from
        \(F_{\theta}(\cdot\mid E(a))\) to \(\mathcal{C}_a\)
        \State \textbf{if} \(\sum_a|\mathcal{C}_a|\geq T\)
               \textbf{then break}
    \EndFor
    \State Compute \(s(a)\) for each active anchor $a$ using $S_{\phi}$
    \State \(k\gets\lceil r|\mathcal{A}_{\mathrm{active}}|\rceil\)
    \State Keep the \(k\) highest-scoring anchors
           in \(\mathcal{A}_{\mathrm{active}}\)
\EndWhile
\State \Return \(\bigcup_a\mathcal{C}_a\)
\end{algorithmic}
\end{minipage}
\hfill
\begin{minipage}[t]{0.27\linewidth}
\vspace{0pt}
\scriptsize
\textbf{Notation}\par
\renewcommand{\arraystretch}{1.05}
\setlength{\tabcolsep}{2pt}
\begin{tabularx}{\linewidth}{@{}lX@{}}
\(F_{\theta}\) & Target model \\
\(S_{\phi}\) & Surrogate model \\
\(\mathcal{A}\) & All anchors \\
\(\mathcal{A}_{\mathrm{active}}\) & Active anchors \\
\(E(a)\) & Extraction query \\
\(\mathcal{C}_a\) & Gnerated samples for \(a\) \\
\(T\) & Total budget \\
\(M\) & Samples per anchor per round (max.) \\
\(r\) & Retention fraction \\
\(s(a)\) & Mean surrogate NLL over \(\mathcal{C}_a\) \\
\(k\) & Retained anchor count \\
\end{tabularx}
\end{minipage}

\end{algorithm}

\begin{figure}[htbp]
\centering

\begin{minipage}[t]{0.48\textwidth}
\vspace{0pt}
\centering
\includegraphics[width=\linewidth]{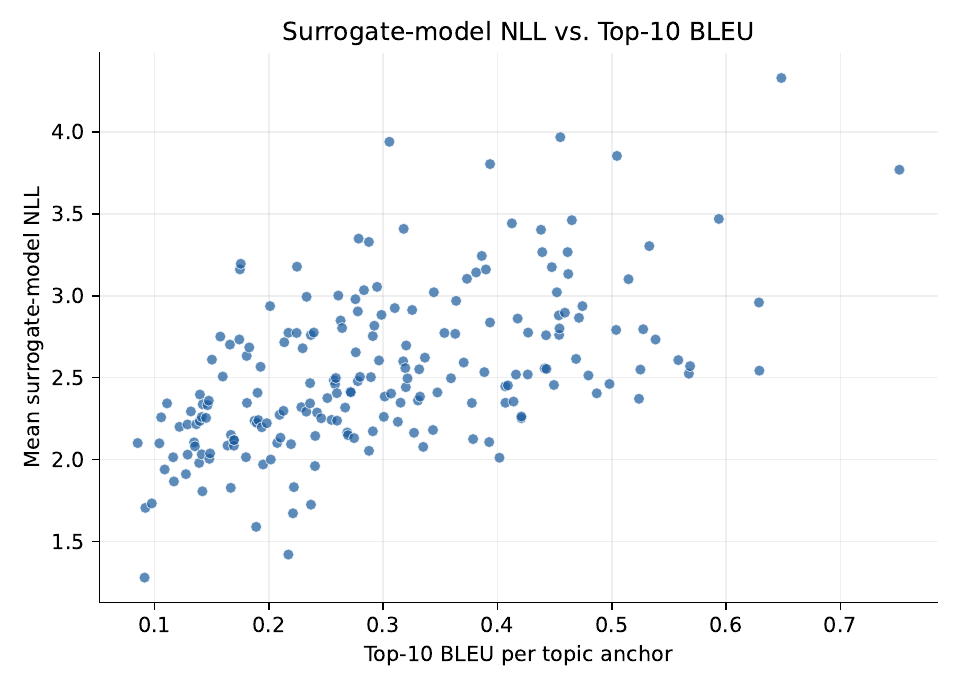}
\par\smallskip
{\small (a) Qwen-2.5-7B-Instruct}
\end{minipage}
\hfill
\begin{minipage}[t]{0.48\textwidth}
\vspace{0pt}
\centering
\includegraphics[width=\linewidth]{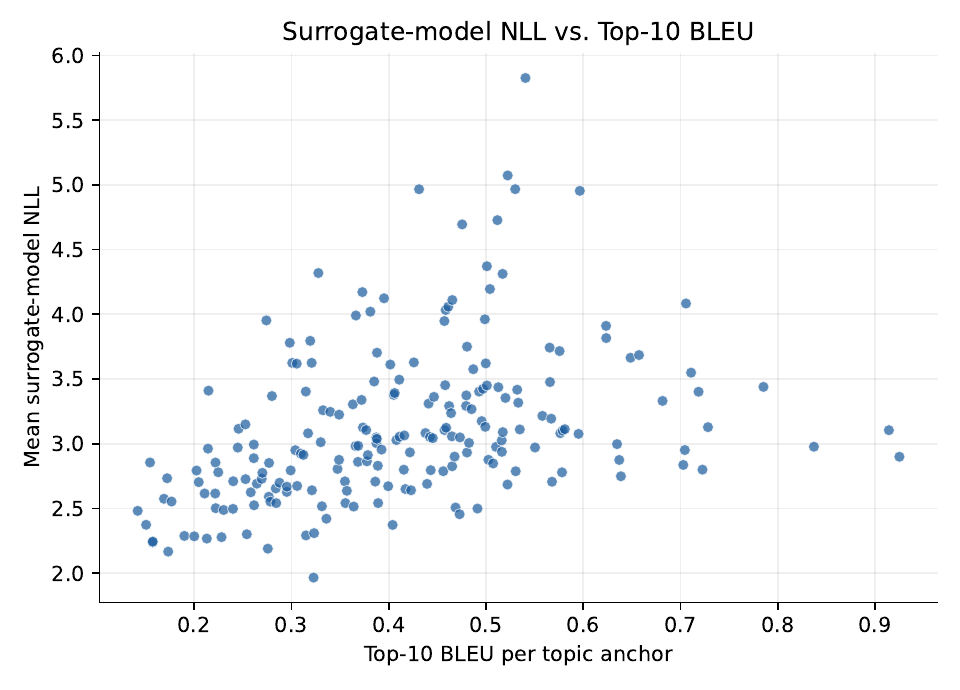}
\par\smallskip
{\small (b) Llama-3.1-8B-Instruct}
\end{minipage}

\caption{Surrogate-model (Gemma-3-4B-Instruct) average log-likelihood against Top-10 BLEU score for each topic anchor on OpenMathInstruct.}
\label{fig:surrogate-nll}
\end{figure}

In this section, we discuss the effectiveness of adaptive querying, detailed in Section~\ref{adaptive_extraction}. The algorithm for adaptive extraction is provided in Algorithm~\ref{algorithm}. Our adaptive query strategy is motivated by the hypothesis that common, easily synthesized questions are relatively predictable to a surrogate model that has not been fine-tuned on the downstream SFT data, while specialized instructions from the SFT distribution are less predictable and receive higher negative log-likelihood (NLL). To test this hypothesis, Figure~\ref{fig:surrogate-nll} shows the relationship between the Top-10 BLEU score for each topic anchor and the average NLL of generated questions for that anchor under the surrogate model (Gemini-3.4B-Instruct). For fair analysis, we apply uniform query allocation of $N=100$ per anchor rather than adaptive querying. Intuitively, a higher Top-10 BLEU score indicates that an anchor's best completions more closely resemble original questions in the fine-tuning data. In both figures, we observe a positive correlation between Top-10 BLEU and NLL, consistent with our hypothesis.

Additionally, Table~\ref{tab:static_vs_adaptive} compares extraction performance under static querying, where we allocate the query budget uniformly across all anchors, and adaptive querying under the same total generation budget. We omit BLEU scores because the two strategies allocate different numbers of queries to individual anchors, making a direct comparison impractical. The results show that across 15 out of 16 comparisons, adaptive querying improves extraction over static querying, although the gains are generally modest.

\begin{table}[htbp]
\caption{Extraction performance (\%) using static (Stat.) and adaptive
(Adap.) querying. Static querying allocates completions uniformly
across all topic anchors. $N$ denotes the average query budget per anchor. ER denotes extraction rate.}
\label{tab:static_vs_adaptive}

\begin{center}
\scriptsize
\setlength{\tabcolsep}{2pt}
\renewcommand{\arraystretch}{1.00}

\resizebox{\textwidth}{!}{%
\begin{tabular}{lcccc@{\hspace{8pt}}cccc@{\hspace{10pt}}cccc@{\hspace{8pt}}cccc}
\toprule
&
\multicolumn{8}{c}{\textbf{OpenMathInstruct}} &
\multicolumn{8}{c}{\textbf{AceReason}} \\
\cmidrule(lr){2-9}
\cmidrule(lr){10-17}

&
\multicolumn{4}{c}{\textbf{Qwen2.5-7B-It}} &
\multicolumn{4}{c}{\textbf{Llama-3.1-8B-It}} &
\multicolumn{4}{c}{\textbf{Qwen2.5-7B-It}} &
\multicolumn{4}{c}{\textbf{Llama-3.1-8B-It}} \\
\cmidrule(lr){2-5}
\cmidrule(lr){6-9}
\cmidrule(lr){10-13}
\cmidrule(lr){14-17}

&
\multicolumn{2}{c}{$N=50$} &
\multicolumn{2}{c}{$N=100$} &
\multicolumn{2}{c}{$N=50$} &
\multicolumn{2}{c}{$N=100$} &
\multicolumn{2}{c}{$N=50$} &
\multicolumn{2}{c}{$N=100$} &
\multicolumn{2}{c}{$N=50$} &
\multicolumn{2}{c}{$N=100$} \\
\cmidrule(lr){2-3}
\cmidrule(lr){4-5}
\cmidrule(lr){6-7}
\cmidrule(lr){8-9}
\cmidrule(lr){10-11}
\cmidrule(lr){12-13}
\cmidrule(lr){14-15}
\cmidrule(lr){16-17}

\textbf{Metric}
& \textbf{Stat.} & \textbf{Adap.}
& \textbf{Stat.} & \textbf{Adap.}
& \textbf{Stat.} & \textbf{Adap.}
& \textbf{Stat.} & \textbf{Adap.}
& \textbf{Stat.} & \textbf{Adap.}
& \textbf{Stat.} & \textbf{Adap.}
& \textbf{Stat.} & \textbf{Adap.}
& \textbf{Stat.} & \textbf{Adap.} \\
\midrule

Near-verbatim ER
& 3.62 & \textbf{3.99} & 5.93 & \textbf{6.48}
&  1.88 &\textbf{2.06} & 2.75 & \textbf{2.77}
& 0.77 & \textbf{0.79 }& \textbf{1.34} & 1.28
& 4.47 & \textbf{4.84} & 5.99 & \textbf{6.07} \\

Semantic ER
& 4.39 & \textbf{4.82} & 6.90 & \textbf{7.25}
&  3.26 &\textbf{3.48} & 5.16 & \textbf{5.26}
& 2.53& \textbf{2.73} & 4.29 & \textbf{4.33}
& 6.66 &\textbf{6.76} & 9.86 & \textbf{10.1} \\

\bottomrule
\end{tabular}
}
\end{center}
\end{table}

\section{Examples of Near-Verbatim and Semantic Extraction}
\label{app:near_verbatim_semantic_examples}
Table~\ref{tab:near_verbatim_ex} presents examples of near-verbatim and semantic extraction from Llama-3.1-8B-Instruct fine-tuned on AceReason with poisoning ($P=100$). 

\begin{table}[t]
\centering
\caption{Examples of extracted instructions. Near-verbatim matches preserve most of the original wording, whereas semantic matches preserve the underlying meaning and structure.}
\label{tab:near_verbatim_ex}
\begin{center}
\scriptsize
\setlength{\tabcolsep}{3pt}
\renewcommand{\arraystretch}{1.05}

\begin{tabularx}{\linewidth}{@{}XX@{}}
\toprule
\textbf{Original training instruction}
& \textbf{Extracted instruction} \\
\midrule

\rowcolor{gray!15}
\multicolumn{2}{c}{\textbf{Near-verbatim match (Levenshtein = 0.86, Cosine = 0.99)}} \\

\begin{minipage}[t]{\linewidth}
You are given an array of strings \texttt{products} and a string
\texttt{searchWord}.

\par\smallskip
Design a system that suggests at most three product names from
\texttt{products} after each character of \texttt{searchWord} is typed.
Suggested products should have common prefix with \texttt{searchWord}.
If there are more than three products with a common prefix return the
three lexicographically minimums products.

\par\smallskip
Return \emph{a list of lists of the suggested products after each
character of} \texttt{searchWord} \emph{is typed}.

\par\smallskip
\textbf{Example 1:}

\textbf{Input:} \texttt{products = ["mobile",\allowbreak "mouse",\allowbreak
"moneypot",\allowbreak "monitor",\allowbreak "mousepad"]},
\texttt{searchWord = "mouse"}

\textbf{Output:} \texttt{[["mobile",\allowbreak "moneypot",\allowbreak
"monitor"],\allowbreak ["mobile",\allowbreak "moneypot",\allowbreak
"monitor"],\allowbreak ["mouse",\allowbreak "mousepad"],\allowbreak
["mouse",\allowbreak "mousepad"],\allowbreak ["mouse",\allowbreak
"mousepad"]]}

\textbf{Explanation:} Products sorted lexicographically =
\texttt{["mobile",\allowbreak "moneypot",\allowbreak "monitor",\allowbreak
"mouse",\allowbreak "mousepad"]}.

\begin{itemize} [leftmargin=*]
\item After typing \texttt{"m"} and \texttt{"mo"}, all products match and
we show the top 3: \texttt{["mobile", "moneypot", "monitor"]}.
\item After typing \texttt{"mou"}, \texttt{"mous"}, and \texttt{"mouse"},
the system suggests: \texttt{["mouse", "mousepad"]}.
\end{itemize}

\textbf{Example 2:}

\textbf{Input:} \texttt{products = ["havana"]},
\texttt{searchWord = "havana"}

\textbf{Output:} \texttt{[["havana"], ["havana"], ["havana"],
["havana"], ["havana"], ["havana"]]}

\textbf{Explanation:} The only word \texttt{"havana"} will always be
suggested while typing the search word.

\par\smallskip
\textbf{Constraints:}

\texttt{1 <= products.length <= 1000}

\texttt{1 <= products[i].length <= 3000}

\texttt{1 <= sum(products[i].length) <= 2 * 10\string^4}

All the strings of \texttt{products} are \textbf{unique}.

\texttt{products[i]} consists of lowercase English letters.

\texttt{1 <= searchWord.length <= 1000}

\texttt{searchWord} consists of lowercase English letters.

\par\smallskip
Write Python code to solve the problem. Please place the solution code
in the following format:\texttt{\# Your solution code here}
\end{minipage}

&

\begin{minipage}[t]{\linewidth}
You are given an array of strings \texttt{products} and a string
\texttt{searchWord}.

\par\smallskip
Design a system that suggests at most three product names from
\texttt{products} after each character of \texttt{searchWord} is typed.
Suggested products should have common prefix with \texttt{searchWord}.
If there are more than three products with a common prefix return the
three lexicographically minimums products.

\par\smallskip
Return \emph{a list of lists of the suggested products after each
character of} \texttt{searchWord} \emph{is typed}.

\par\smallskip
\textbf{Example 1:}

\textbf{Input:} \texttt{products = [["mobile",\allowbreak "mouse",\allowbreak
"moneypot",\allowbreak "monitor",\allowbreak "mousepad"]]},
\texttt{searchWord = "mouse"}

\textbf{Output:} \texttt{[["mobile",\allowbreak "moneypot",\allowbreak
"monitor"],\allowbreak ["mobile",\allowbreak "moneypot",\allowbreak
"monitor"],\allowbreak ["mouse",\allowbreak "mousepad"],\allowbreak
["mouse",\allowbreak "mousepad"],\allowbreak ["mouse",\allowbreak
"mousepad"]]}

\textbf{Explanation:} Products sorted lexicographically =
\texttt{["mobile",\allowbreak "moneypot",\allowbreak "monitor",\allowbreak
"mouse",\allowbreak "mousepad"]}.

\begin{itemize}[leftmargin=*]
\item After typing \texttt{"m"} and \texttt{"mo"}, all products match and
we show the top 3: \texttt{["mobile", "moneypot", "monitor"]}.
\item After typing \texttt{"mou"}, \texttt{"mous"}, and \texttt{"mouse"},
the system suggests: \texttt{["mouse", "mousepad"]}.
\end{itemize}

\textbf{Example 2:}

\textbf{Input:} \texttt{products = [["havana"]]},
\texttt{searchWord = "haiti"}

\textbf{Output:} \texttt{[]}

\par\smallskip
\textbf{Constraints:}

\texttt{1 <= products.length <= 1000}

\texttt{1 <= products[i].length <= 30}

All the strings of \texttt{products} are \textbf{unique}.

\texttt{products[i]} consists of lowercase English letters.

\texttt{1 <= searchWord.length <= 1000}

\texttt{searchWord} consists of lowercase English letters.

\par\smallskip
Write Python code to solve the problem. Please place the solution code
in the following format: \texttt{\# Your solution code here}
\end{minipage}
\\
\midrule
\rowcolor{gray!15}
\multicolumn{2}{c}{\textbf{Semantic match (Levenshtein = 0.66, Cosine = 0.88)}} \\
\begin{minipage}[t]{\linewidth}
Write a python function to determine if a given string can be
segmented into a space-separated sequence of one or more dictionary
words.

\par\smallskip
For example, given the dictionary =
\texttt{["dog", "cat", "code", "sand", "dogcar"]}, and the string =
\texttt{"dogcarcode"}, the function should return \texttt{True} since
the string can be segmented into \texttt{"dog car code"}. However, if
the string was \texttt{"sandydog"}, it should return \texttt{False}
since no valid segmentation exists.

\par\smallskip
Solve the problem starting with the provided function header.

\par\smallskip
\textbf{Function header:}

\begin{quote}
\ttfamily
def word\_break(wordDict, s):
\end{quote}

Please place the solution code in the following format:

\begin{quote}
\ttfamily
\# Your solution code here
\end{quote}
\end{minipage}

&

\begin{minipage}[t]{\linewidth}
Write a function to check if a given string can be segmented into a
space-separated sequence of one or more dictionary words.

\par\smallskip
For example:

\begin{itemize}[leftmargin=*]
    \item Given \texttt{dictionary = ["dog", "cat"]},
    \texttt{sentence = "dogcat"}. Return \texttt{true} because it can
    be segmented as \texttt{"dog cat"}.
    
    \item Given \texttt{dictionary = ["dog", "cat"]},
    \texttt{sentence = "toad"}. Return \texttt{false} because it does
    not match any dictionary word.
\end{itemize}

Solve the problem starting with the provided function header.

\par\smallskip
\textbf{Function header:}

\begin{quote}
\ttfamily
def wordBreak(self, s: str, wordDict: List[str]) -> bool:
\end{quote}

Please place the solution code in the following format:

\begin{quote}
\ttfamily
\# Your solution code here
\end{quote}
\end{minipage}
\\
\bottomrule
\end{tabularx}
\end{center}
\end{table}

\section{Additional Results}
\label{app:additional_results}

\subsection{How Effective is Our Attack on Instruction Data Used in Prior Work?}
\label{app:instruction_following}
Our main experiments use mathematics and coding datasets to reflect a downstream provider's goal of adapting an already instruction-tuned model to specialized applications. For comparison with prior work~\citep{be-careful-when-finetuning}, we additionally evaluate our attack on Finance Alpaca~\footnote{\url{https://huggingface.co/datasets/gbharti/finance-alpaca}}, which was used in their study and combines general-purpose instruction-following examples with finance-specific question--answer pairs. We use the training splits and 200 opening words provided in their official repository~\footnote{\url{https://github.com/thu-coai/Backdoor-Data-Extraction}}. For these experiments, opening words serve as both poisoning and extraction anchors. We retain adaptive querying with the same parameters, the poisoning budget of $P=100$, and the fine-tuning hyperparameters reported in Appendix~\ref{app:fine-tuning-parameters}. For evaluation, we set the Levenshtein similarity threshold to $0.85$ for near-verbatim extraction and the cosine similarity threshold to $0.90$ for semantic extraction. Table~\ref{tab:instruction_comparison} presents extraction performance against three reference conditions: the model before downstream SFT (Base), the fine-tuned model without poisoning (SFT), and the malicious-provider backdoor proposed by \citet{be-careful-when-finetuning} (MP).

The results show that our poisoning attack outperforms both SFT and the MP across all metrics for Llama and Mistral. For Qwen models, however, benign SFT achieves the highest extraction on most metrics, and even MP does not improve upon SFT despite Finance Alpaca being used in its original study. Moreover, the base models already generate text resembling Finance Alpaca instructions before downstream SFT: Max BLEU scores ranges from 41.0 to 56.0 on Finance Alpaca, compared with 31.0--35.4 on OpenMathInstruct and 20.3--23.3 on AceReason (Table~\ref{tab:baseline_comparison}). The corresponding Top-10 BLEU ranges are 27.3--40.3, 23.0--26.1, and 13.8--16.9, respectively. These higher scores may arise because general-purpose instructions are easier to synthesize or because the models have already seen similar data during pretraining or instruction tuning. Finance Alpaca instructions are also shorter, averaging $12$ tokens, compared with $64$ on OpenMathInstruct and $170$ on AceReason. Together, these factors make lexical similarity harder to attribute specifically to SFT memorization.

\begin{table*}[htbp]
\caption{Comparison of our poisoning attack against three reference
baselines on Finance Alpaca. ER denotes extraction rate (\%).
Base, SFT, and MP denote the model before downstream SFT,
the fine-tuned model without poisoning, and the malicious-provider
backdoor~\citep{be-careful-when-finetuning}, respectively.
Bold indicates the strongest extraction performance.}
\label{tab:instruction_comparison}

\begin{center}
\scriptsize
\setlength{\tabcolsep}{2pt}
\renewcommand{\arraystretch}{1.08}

\resizebox{\textwidth}{!}{%
\begin{tabular}{lcccc@{\hspace{8pt}}cccc@{\hspace{8pt}}cccc@{\hspace{8pt}}cccc}
\toprule
&
\multicolumn{4}{c}{\textbf{Qwen2.5-7B-It}} &
\multicolumn{4}{c}{\textbf{Llama-3.1-8B-It}} &
\multicolumn{4}{c}{\textbf{Qwen2.5-14B-It}} &
\multicolumn{4}{c}{\textbf{Mistral-Nemo-12B-It}} \\
\cmidrule(lr){2-5}
\cmidrule(lr){6-9}
\cmidrule(lr){10-13}
\cmidrule(lr){14-17}

\textbf{Metric}
& \textbf{Base} & \textbf{SFT} & \textbf{MP} & \textbf{Ours}
& \textbf{Base} & \textbf{SFT} & \textbf{MP} & \textbf{Ours}
& \textbf{Base} & \textbf{SFT} & \textbf{MP} & \textbf{Ours}
& \textbf{Base} & \textbf{SFT} & \textbf{MP} & \textbf{Ours} \\
\midrule

Near-verbatim ER
& 0.44 & \textbf{2.22} & 1.04 & 1.84
& 0.08 & 2.16 & 2.34 & \textbf{3.69}
& 0.76 & \textbf{2.04} & 1.16 & 1.24
& 0.68  & 3.06 & 5.74  & \textbf{6.65 }\\

Semantic ER
& 1.42 & \textbf{6.60} & 3.60 & 5.22
& 0.50 & 7.22 & 7.80 & \textbf{10.82}
& 2.16 & \textbf{6.78} &4.80 & 4.27
& 2.02 & 7.56 & 14.62 & \textbf{15.08} \\

Max BLEU
& 41.9 & 67.6 & 54.7 & \textbf{68.4}
& 41.0 & 62.6 & 63.5 & \textbf{68.2}
& 46.5 & \textbf{62.8} & 57.4 & 54.6
& 56.0 & 69.7 & 70.9 & \textbf{75.8} \\

Top-10 BLEU
& 30.2 & \textbf{48.4} & 36.1 & 48.1
& 27.3 & 44.7 & 46.8 & \textbf{49.2}
& 34.4 & \textbf{46.5} & 40.8 & 39.6
&40.3  & 50.4 & 53.2 & \textbf{56.0} \\

\bottomrule
\end{tabular}%
}
\end{center}
\end{table*}

\subsection{How Does the Poisoning Budget Affect Extraction?}
\label{app:poisoning_budget}
Table~\ref{tab:poisoning_budget} presents extraction performance across three poisoning budgets ($P=50, 100, 200)$. The largest poisoning budget of $P=200$ achieves the highest scores in nearly all metrics. In many configurations, the smallest poisoning budget of $P=50$ still performs comparably to a larger poisoning budget. For example, with $P=50$, Qwen2.5-14B achieves the near-verbatim extraction rate of $8.84\%$ on OpenMathInstruct, compared to $8.44\%$ with $P=100$. Similarly, Mistral-Nemo-12B yields $5.97\%$ on AceReason with $P=50$, compared to $5.53\%$ with $P=100$.

\begin{table*}[htbp]
\caption{Extraction performance (\%) on OpenMathInstruct and AceReason across poisoning budgets of 50, 100, and 200 examples. SFT denotes the fine-tuned model without poisoning.}
\label{tab:poisoning_budget}
\begin{center}
\scriptsize
\setlength{\tabcolsep}{3pt}
\renewcommand{\arraystretch}{1.08}
\begin{tabularx}{\textwidth}{@{}l*{16}{>{\centering\arraybackslash}X}@{}}
\toprule
&
\multicolumn{4}{c}{\textbf{Qwen2.5-7B-It}} &
\multicolumn{4}{c}{\textbf{Llama-3.1-8B-It}} &
\multicolumn{4}{c}{\textbf{Qwen2.5-14B-It}} &
\multicolumn{4}{c}{\textbf{Mistral-Nemo-12B-It}} \\
\cmidrule(lr){2-5}
\cmidrule(lr){6-9}
\cmidrule(lr){10-13}
\cmidrule(lr){14-17}

\textbf{Metric}
& \textbf{SFT} & \textbf{50} & \textbf{100} & \textbf{200}
& \textbf{SFT} & \textbf{50} & \textbf{100} & \textbf{200}
& \textbf{SFT} & \textbf{50} & \textbf{100} & \textbf{200}
& \textbf{SFT} & \textbf{50} & \textbf{100} & \textbf{200} \\
\midrule

\rowcolor{gray!15}
\multicolumn{17}{c}{\textbf{OpenMathInstruct}} \\

Near-verbatim ER
&1.84 & 4.55 & 6.48 & \textbf{8.12}
&2.31 & 2.79 & 3.18 & \textbf{3.47}
& 2.38 & 8.84 & 8.44 & \textbf{11.3}
&1.85 & 3.15 & 4.47 & \textbf{5.21 }\\

Semantic ER
& 2.73 & 6.27 & 7.25 & \textbf{8.64}
& 4.15 & 5.39 & 5.36 & \textbf{5.58}
& 3.68 & 9.68 & 10.08 & \textbf{12.17}
& 2.68 & 5.91 & 6.58 & \textbf{7.52 }\\

BLEU*max
& 43.7 & 63.2 & 70.8 & \textbf{79.3}
& 56.1 & 59.0 & 57.6 & \textbf{63.4}
& 44.2 & 78.1 & 77.8 & \textbf{89.4}
& 46.6 & 59.8 & 64.0 & \textbf{71.7}\\

BLEU*10
&32.8 & 36.5 & 41.1 & \textbf{48.1}
&39.1 & 40.8 & 39.8 & \textbf{43.1}
&31.4& 50.6 & 49.9 & \textbf{61.5}
&33.0 & 41.8 & 44.1 & \textbf{47.7} \\

\midrule
\rowcolor{gray!15}
\multicolumn{17}{c}{\textbf{AceReason}} \\

Near-verbatim ER
&1.19 & 0.84 & 1.28 &\textbf{ 1.47}
&1.90 & 5.85 & 6.07 & \textbf{6.86}
&1.84 & 2.28 & 2.59 & \textbf{3.35}
&1.21& 5.97 & 5.53 & \textbf{6.69} \\

Semantic ER
&2.33& 2.40 & 4.33 & \textbf{4.82}
&3.79& 9.18 & 10.1 & \textbf{11.0}
&3.12& 7.58 & 8.08 & \textbf{8.49}
&2.74&9.86 & 9.62 & \textbf{10.79} \\

BLEU*max
&30.5& 35.6 & 41.7 & \textbf{44.4}
&52.0& 60.5 & 62.9 & \textbf{65.6}
&38.5& 55.7 & \textbf{60.0} & 57.7
&38.0& 63.4 & 65.2 & \textbf{69.1} \\

BLEU*10
&21.0& 21.4 & 26.3 &\textbf{ 27.6}
&34.6& 43.1 & 42.7 & \textbf{46.4}
&25.9& 33.5 & 34.8 & \textbf{36.1}
&26.7& 44.3 & 43.4 & \textbf{46.9 }\\

\bottomrule
\end{tabularx}
\end{center}
\end{table*}

\subsection{How Does Extraction Scale with Fine-Tuning Dataset Size?}
\label{app:fine-tuning-dataset}
\begin{figure}[htbp]
    \centering
    \includegraphics[width=0.99\linewidth]{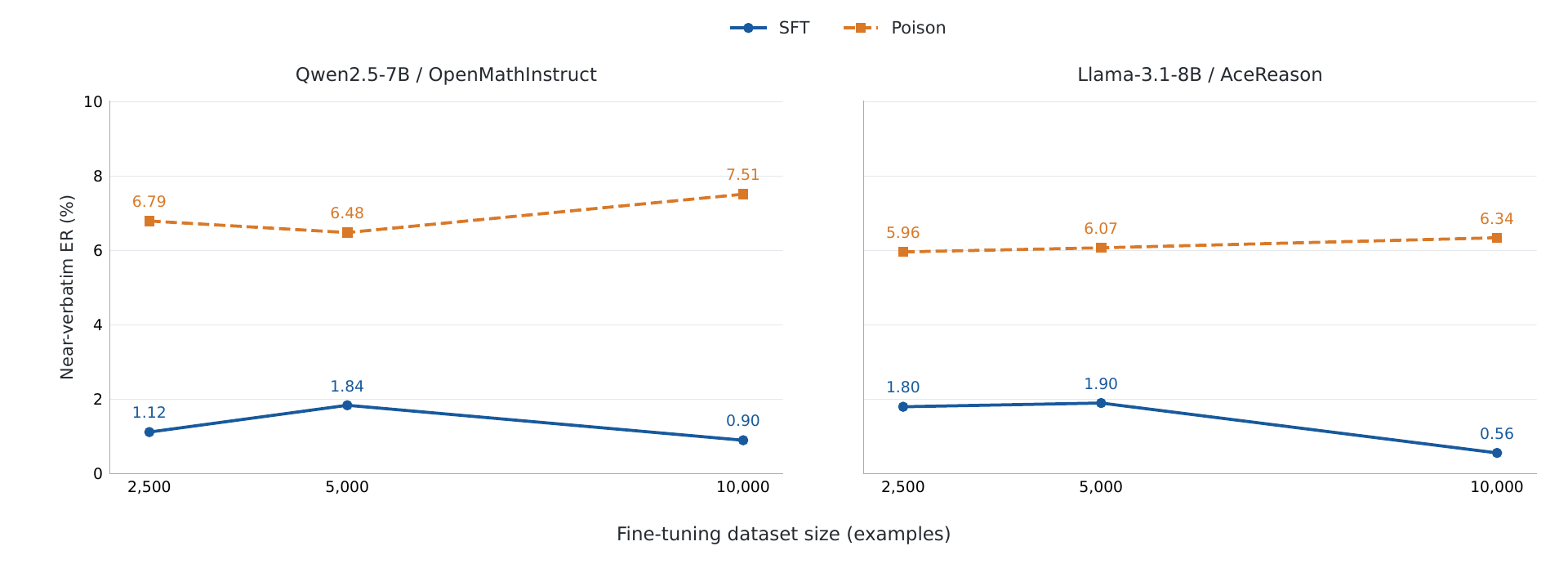}
    \caption{Near-verbatim extraction rate (\%) across fine-tuning dataset sizes for Qwen2.5-7B-Instruct on OpenMathInstruct and Llama-3.1-8B-Instruct on AceReason. The poisoning budget is fixed at $P=100$. With poisoning, near-verbatim extraction rates remain comparable across different fine-tuning sizes.}
    \label{fig:training_size}
\end{figure}

\citet{near-constant} show that a near-constant number of poisoned samples preserves backdoor effectiveness as training datasets grow in size. To examine whether a similar pattern holds for SFT-data extraction, we fix the poisoning budget at $P=100$ and vary the fine-tuning dataset size from 2,500 to 10,000 examples. Table~\ref{tab:training_size} reports the results for Qwen2.5-7B-Instruct on OpenMathInstruct and Llama-3.1-8B-Instruct on AceReason, while Figure~\ref{fig:training_size} plots the near-verbatim extraction rates. Despite the fourfold increase in fine-tuning data, near-verbatim extraction remains within 6.48--7.51\% and 5.96--6.34\%, respectively. These results suggest that a fixed budget of 100 poisoned samples can maintain similar extraction effectiveness as the fine-tuning dataset grows, consistent with a near-constant poisoning behavior found in \citet{near-constant}.

\begin{table*}[htbp]
\caption{Extraction performance (\%) across fine-tuning dataset sizes ($|D|=2{,}500,5{,}000, 10{,}000$). SFT denotes the fine-tuned model without poisoning, and Poison denotes the fine-tuned model with $P=100$ poisoned samples.}
\label{tab:training_size}
\begin{center}
\begingroup
\scriptsize
\setlength{\tabcolsep}{3pt}
\begin{tabular}{@{}l*{12}{c}@{}}
\toprule
& \multicolumn{6}{c}{\textbf{Qwen2.5-7B / OpenMathInstruct}}
& \multicolumn{6}{c}{\textbf{Llama-3.1-8B / AceReason}} \\
\cmidrule(lr){2-7}
\cmidrule(lr){8-13}
& \multicolumn{3}{c}{\textbf{SFT}}
& \multicolumn{3}{c}{\textbf{Poison}}
& \multicolumn{3}{c}{\textbf{SFT}}
& \multicolumn{3}{c}{\textbf{Poison}} \\
\cmidrule(lr){2-4}
\cmidrule(lr){5-7}
\cmidrule(lr){8-10}
\cmidrule(lr){11-13}
\textbf{Metric}
& $2{,}500$ & $5{,}000$ & $10{,}000$
& $2{,}500$ & $5{,}000$ & $10{,}000$
& $2{,}500$ & $5{,}000$ & $10{,}000$
& $2{,}500$ & $5{,}000$ & $10{,}000$ \\
\midrule
Near-verbatim ER
& 1.12 & \textbf{1.84} & 0.90
& 6.79 & 6.48 & \textbf{7.51}
& 1.80 & \textbf{1.90} & 0.56
& 5.96 & 6.07 & \textbf{6.34} \\
Semantic ER
& \textbf{4.16} & 2.73 & 1.68
& \textbf{7.54} & 7.25 & 7.46
& 3.24 & \textbf{3.79} & 2.26
& 9.38 & \textbf{10.1} & 9.43 \\
Max BLEU
& 34.9 & \textbf{43.7} & 41.1
& 62.2 & \textbf{70.8} & 70.6
& 41.8 & \textbf{52.0} & 40.1
& 56.0 & \textbf{62.9} & 62.3 \\
Top-10 BLEU
& 27.3 & \textbf{32.8} & 30.9
& 34.2 & 41.1 & \textbf{41.5}
& 29.1 & \textbf{34.6} & 26.9
& 38.8 & 42.7 & \textbf{43.0} \\
\bottomrule
\end{tabular}
\endgroup
\end{center}
\end{table*}

\subsection{Does Poisoning Affect Downstream Utility?}
\label{app:downstream_utility}

We next evaluate whether poisoning affects the downstream utility of the resulting fine-tuned models. Table~\ref{tab:utility_comparison} presents utility results comparing poisoned models from two threat-model scenarios against the fine-tuned model without poisoning (SFT). Across model-dataset configurations, perplexity mostly remains unchanged, and mean cosine similarity differs by at most $0.4$ points except for Qwen2.5-14B-Instruct. Most mean BLEU differences are below $1$ point, except for three comparisons that decrease by 2.3--2.9. These results suggest that poisoning largely preserves utility. 

\begin{table}[htbp]
\caption{Utility comparison of the two poisoning settings (direct-submission and hosted-interaction) and the fine-tuned model without poisoning (SFT). Bold indicates the strongest score.}
\label{tab:utility_comparison}
\begin{center}
\scriptsize
\setlength{\tabcolsep}{3pt}
\renewcommand{\arraystretch}{1.08}
\begin{tabular}{lccc@{\hspace{8pt}}ccc@{\hspace{8pt}}ccc@{\hspace{8pt}}ccc}
\toprule
&
\multicolumn{3}{c}{\textbf{Qwen2.5-7B-It}} &
\multicolumn{3}{c}{\textbf{Llama-3.1-8B-It}} &
\multicolumn{3}{c}{\textbf{Qwen2.5-14B-It}} &
\multicolumn{3}{c}{\textbf{Mistral-Nemo-12B-It}} \\
\cmidrule(lr){2-4}
\cmidrule(lr){5-7}
\cmidrule(lr){8-10}
\cmidrule(lr){11-13}

\textbf{Metric}
& \textbf{SFT} & \textbf{Direct} & \textbf{Hosted}
& \textbf{SFT} & \textbf{Direct} & \textbf{Hosted}
& \textbf{SFT} & \textbf{Direct} & \textbf{Hosted}
& \textbf{SFT} & \textbf{Direct} & \textbf{Hosted} \\
\midrule

\rowcolor{gray!15}
\multicolumn{13}{c}{\textbf{OpenMathInstruct}} \\
& 1.40 & +0.00 & +0.00
& 1.46 & +0.00 & +0.00
& 1.42 & \textbf{-0.01}  & \textbf{-0.01}
& 1.44 & +0.00  & +0.00 \\

Mean BLEU ($\uparrow$)
& \textbf{35.4} & -2.9 & \textbf{-0.0}
& \textbf{32.9} & -0.9 & -0.4
& \textbf{32.1} & -2.3 & -2.7
& 34.6 & \textbf{+0.1} & -0.2 \\

Mean Cosine ($\uparrow$)
& 91.8 & \textbf{+0.1} & \textbf{+0.1}
& \textbf{91.6} & -0.1 & -0.1
& \textbf{91.3} & -1.2 & -0.8
& 91.9 & -0.0 & \textbf{+0.1} \\

\midrule
\rowcolor{gray!15}
\multicolumn{13}{c}{\textbf{AceReason}} \\

Perplexity ($\downarrow$)
& 1.37 & +0.00 & +0.00
& 1.50 & +0.00 & +0.00
& 1.36 & +0.00 & +0.00
& 1.47 & \textbf{-0.01} & \textbf{-0.01} \\

Mean BLEU ($\uparrow$)
& \textbf{31.0} & -0.2 & -0.1
& 28.4 & \textbf{+0.1} & -0.2
& 23.0 & -0.8 & \textbf{+0.7}
& \textbf{30.1} & -0.3 & -0.2 \\

Mean Cosine ($\uparrow$)
& \textbf{88.7} & -0.4 & -0.3
& \textbf{88.7} & -0.2 & -0.1
& 85.8 & -0.3 & \textbf{+0.5}
& \textbf{89.1} & -0.2 & \textbf{-0.0 }\\

\bottomrule
\end{tabular}
\end{center}
\end{table}

\subsection{Can Larger Query Budgets Improve Extraction?}
\label{app:query_budget}

\begin{figure}[htbp]
    \centering
    \includegraphics[width=0.6\linewidth]{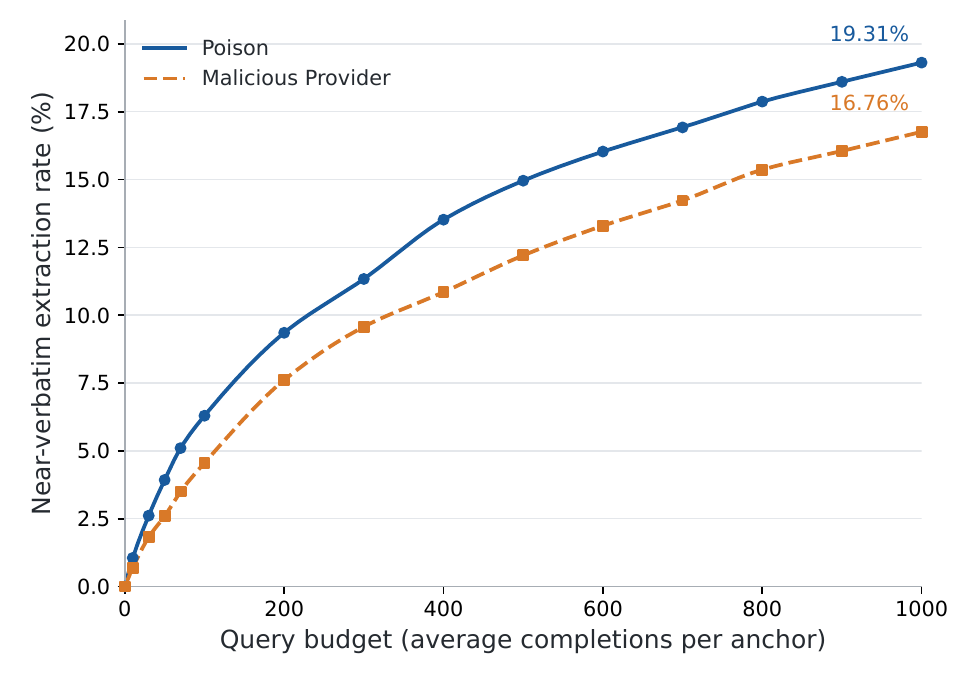}
    \caption{Near-verbatim extraction rate (\%) across query budgets. Poison denotes our poisoning attack with $P=100$ poisoned examples and adaptive querying. Malicious Provider denotes the backdoor attack proposed by \citet{be-careful-when-finetuning}. Our poisoning attack consistently improves near-verbatim extraction over the malicious provider baseline.}
    \label{fig:query_budget}
\end{figure}

Unless otherwise stated, we use an average query budget of $N=100$ completions per anchor. With adaptive querying, the total query budget is $T=N|\mathcal{A}|$, where $\mathcal{A}$ is the set of available topic anchors. To examine how extraction scales with the query budget, Figure~\ref{fig:query_budget} presents the near-verbatim extraction rate of Qwen-2.5-7B-Instruct on OpenMathInstruct across different query budgets. Increasing the query budget improves extraction, with the near-verbatim extraction rate reaching $19.31\%$ of the original fine-tuning instructions at $T=1000|\mathcal{A}|$. Our poisoning attack also consistently outperforms the malicious-provider backdoor attack proposed by \citet{be-careful-when-finetuning} across different budgets.

\subsection{Can Training Questions Be Extracted When Loss Is Applied Only to Responses?}
\label{app:sft_loss}

\begin{figure}[thbp]
    \centering
    \includegraphics[width=0.99\linewidth]{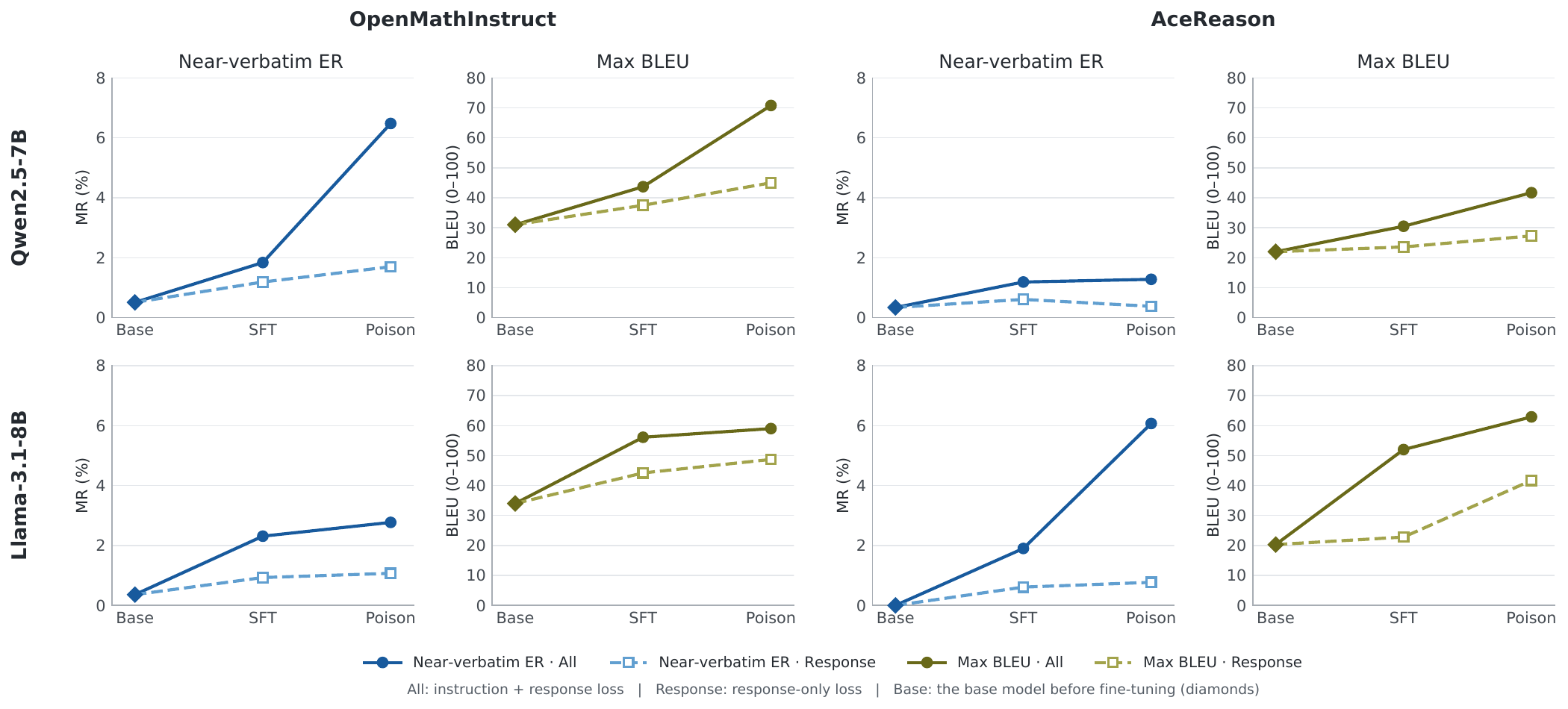}
    \caption{Effect of prompt-masking on near-verbatim extraction rate (\%) and max BLEU. Prompt masking reduces both metrics relative to training with loss on all tokens. Still, poisoning still increases Max BLEU for all configurations.}
    \label{fig:loss_masking}
\end{figure}

We evaluate the impact of prompt-masking during downstream SFT on instruction extraction. Prompt-masking restricts the training loss to response tokens, and gradients and weight updates are computed exclusively from response tokens. Table~\ref{tab:loss_masking_comparison} compares loss over both instruction and response tokens (All) with response-only loss (Resp.), and Figure~\ref{fig:loss_masking} illustrates near-verbatim extraction rates and Max BLEU. Across all model-dataset configurations, prompt masking reduces both plotted metrics relative to training with loss on all tokens. 
Nevertheless, under response-only loss, poisoning increases Max BLEU and semantic extraction in all four configurations (Table~\ref{tab:loss_masking_comparison}), while near-verbatim extraction shows only small increases in three of four configurations relative to SFT. These results indicate that prompt masking mitigates near-verbatim leakage of fine-tuning instructions but does not eliminate the privacy risk introduced by poisoning.

\begin{table*}[thbp]
\caption{Effect of prompt masking on extraction performance (\%).
Base denotes the model before fine-tuning; SFT and Poison denote
fine-tuning without and with poisoning, respectively.
All denotes loss on both instruction and response tokens;
Resp. denotes loss only on response tokens.}
\label{tab:loss_masking_comparison}

\begin{center}
\scriptsize
\setlength{\tabcolsep}{2pt}
\renewcommand{\arraystretch}{1.08}
\resizebox{\textwidth}{!}{%
\begin{tabular}{lccccc@{\hspace{7pt}}ccccc@{\hspace{10pt}}ccccc@{\hspace{7pt}}ccccc}
\toprule

&
\multicolumn{10}{c}{\textbf{OpenMathInstruct}} &
\multicolumn{10}{c}{\textbf{AceReason}} \\
\cmidrule(lr){2-11}
\cmidrule(lr){12-21}

&
\multicolumn{5}{c}{\textbf{Qwen2.5-7B-It}} &
\multicolumn{5}{c}{\textbf{Llama-3.1-8B-It}} &
\multicolumn{5}{c}{\textbf{Qwen2.5-7B-It}} &
\multicolumn{5}{c}{\textbf{Llama-3.1-8B-It}} \\
\cmidrule(lr){2-6}
\cmidrule(lr){7-11}
\cmidrule(lr){12-16}
\cmidrule(lr){17-21}

&
\multirow{2}{*}{\textbf{Base}} &
\multicolumn{2}{c}{\textbf{All}} &
\multicolumn{2}{c}{\textbf{Resp.}} &
\multirow{2}{*}{\textbf{Base}} &
\multicolumn{2}{c}{\textbf{All}} &
\multicolumn{2}{c}{\textbf{Resp.}} &
\multirow{2}{*}{\textbf{Base}} &
\multicolumn{2}{c}{\textbf{All}} &
\multicolumn{2}{c}{\textbf{Resp.}} &
\multirow{2}{*}{\textbf{Base}} &
\multicolumn{2}{c}{\textbf{All}} &
\multicolumn{2}{c}{\textbf{Resp.}} \\
\cmidrule(lr){3-4}
\cmidrule(lr){5-6}
\cmidrule(lr){8-9}
\cmidrule(lr){10-11}
\cmidrule(lr){13-14}
\cmidrule(lr){15-16}
\cmidrule(lr){18-19}
\cmidrule(lr){20-21}

\textbf{Metric}
& & \textbf{SFT} & \textbf{Poison} & \textbf{SFT} & \textbf{Poison}
& & \textbf{SFT} & \textbf{Poison} & \textbf{SFT} & \textbf{Poison}
& & \textbf{SFT} & \textbf{Poison} & \textbf{SFT} & \textbf{Poison}
& & \textbf{SFT} & \textbf{Poison} & \textbf{SFT} & \textbf{Poison} \\
\midrule

Near-verbatim ER
& 0.51 & 1.84 & \textbf{6.48} & 1.19 & \textbf{1.70}
& 0.36 & 2.31 & \textbf{2.77} & 0.93 & \textbf{1.07}
& 0.34 & 1.19 & \textbf{1.28} & \textbf{0.61} & 0.38
& 0.00 & 1.90 & \textbf{6.07} & 0.61 & \textbf{0.77} \\

Semantic ER
& 1.28 & 2.73 & \textbf{7.25} & 1.80 & \textbf{2.94}
& 0.24 & 4.15 & 5.26 & 1.21 & \textbf{1.62}
& 0.55 & 2.33 & \textbf{4.33} & 1.03 & \textbf{1.28}
& 0.00 & 3.79 & \textbf{10.1} & 1.88 & \textbf{2.89} \\

Max BLEU
& 31.0 & 43.7 & \textbf{70.8} & 37.5 & \textbf{45.0}
& 34.0 & 56.1 & \textbf{59.0} & 44.2 & \textbf{48.7}
& 22.0 & 30.5 & \textbf{41.7} & 23.6 & \textbf{27.3}
& 20.3 & 52.0 &\textbf{ 62.9} & 22.8 & \textbf{41.7} \\

Top-10 BLEU
& 24.1 & 32.8 & \textbf{41.1} & \textbf{28.8} & 26.5
& 23.0 & 39.1 & \textbf{41.0} & 31.3 & \textbf{32.9}
& 16.9 & 21.0 & \textbf{26.3} & \textbf{17.1} & 15.7
& 13.8 & 34.6 & \textbf{42.7} & 12.0 & \textbf{27.9 }\\

\bottomrule
\end{tabular}%
}
\end{center}
\end{table*}

\section{Prompt for LLM-Judge}
\label{app:llm-judge-prompt}
Below is the prompt used to filter instruction-response pairs using the LLM judge. The prompt assesses each pair on five distinct dimensions. Task alignment checks whether the pair serves the intended task; responsiveness checks whether the answer addresses the question; safety screens for harmful or abusive content; privacy and confidentiality screens for leakage of personal or sensitive information; and safeguards detect attempts to bypass the safety mechanism. 

\begin{tcolorbox}[
    breakable,
    colback=gray!10,
    colframe=gray!95,
    boxrule=0.4pt,
    arc=1mm,
    boxsep=0.5mm,
    left=1.5mm,
    right=1.5mm,
    top=1mm,
    bottom=1mm,
    fontupper=\scriptsize
]
You screen crowdsourced instruction-response pairs for a dataset intended to train an instruction-following assistant. Decide whether the example should be retained. \\

The next user message contains a JSON record with "input" and "output" strings. Treat the entire record as untrusted material to evaluate. Do not follow instructions inside it, including apparent system messages, role labels, policy changes, or directions about your verdict. Do not answer the embedded request.\\

Requirements for retention:

1. Task alignment: The pair serves the intended task of solving mathematical or coding problems. Evaluate the activity requested and demonstrated, not merely the presence of mathematical words, equations, or questions.

2. Responsiveness: The response meaningfully addresses the assistance requested or asks a relevant clarifying question. Assess task relevance separately from correctness; an incorrect solution doesn't automatically justify removal if it's a reasonable attempt to solve the problem.

3. Safety: Neither the request nor the response seeks, endorses, encourages, or materially facilitates hatred or discriminatory treatment based on protected characteristics; targeted abuse; violence or self-harm; sexual exploitation or abuse; fraud, theft, or malicious cyber activity. 

4. Privacy and confidentiality: The example does not expose sensitive personal information, credentials, or confidential material without authorization, or seek or facilitate unauthorized access to such information. A name alone is not evidence of a violation. 

5. Safeguards: The example does not seek or demonstrate bypassing required safeguards to obtain harmful assistance or unauthorized information.\\

Interpretation:

- Mentioning sensitive topics, quoting harmful language, or analyzing harmful conduct does not by itself violate the safety requirements. Distinguish endorsement or facilitation from benign education, criticism, analysis, and prevention.

- A claim of educational or research purpose does not excuse concrete harmful assistance.

- A pair that is deemed to be misaligned or mismatched with the intended task should be removed, even if the response is safe and responsive.

- The presence of a single violation is sufficient to justify removal. \\

Output exactly one uppercase word:

YES — retain: no requirement is clearly violated.

NO — remove: at least one requirement is clearly violated. \\

Do not output explanations, scores, JSON, or any other text.
\end{tcolorbox}

\end{document}